\documentclass[11pt,a4paper]{article}
\pdfoutput=1
\usepackage{jcappub}
\allowdisplaybreaks[1]
\usepackage{graphicx}
\usepackage[english]{babel}
\usepackage[utf8]{inputenc}
\usepackage{setspace}
\usepackage{url}
\usepackage{verbatim}
\numberwithin{equation}{section}

\newcommand{\ee}{\text{e}}
\renewcommand\({\left(}
\renewcommand\){\right)}
\renewcommand\[{\left[}

\newcommand{\CLASS}{{\sc class}}

\def \and {\qquad\mathrm{and}\qquad}

\def \be {\begin{equation}}
\def \ee {\end{equation}}
\def \bea {\begin{eqnarray}}
\def \eea {\end{eqnarray}}

\def \mr {\mathrm}

\begin{document}
\mbox{}\\
\hfill CERN-PH-TH/2013-313, LAPTH-073/13
\title{The effective gravitational decoupling between dark matter and the CMB}
\author[a]{Luc Voruz,}
\author[a,b,c]{Julien Lesgourgues,}
\author[a]{and Thomas Tram}
\affiliation[a]{Institut de
Th\'eorie des Ph\'enom\`enes Physiques, \'Ecole Polytechnique
F\'ed\'erale de Lausanne, CH-1015, Lausanne,
Switzerland}
\affiliation[b]{CERN, Theory Division, CH-1211 Geneva 23, Switzerland}
\affiliation[c]{LAPTh (CNRS - Universit\'e de Savoie), BP 110, F-74941 Annecy-le-Vieux Cedex, France}
\emailAdd{luc.voruz@gmail.com}
\emailAdd{Julien.Lesgourgues@cern.ch}
\emailAdd{thomas.tram@epfl.ch}
\abstract{We present a detailed and self-contained analytical derivation of the evolution of sub-horizon cosmological perturbations before decoupling, based on previous work by S. Weinberg. These solutions are valid in the minimal $\Lambda$CDM scenario, to first order in perturbation theory, in the tight-coupling limit and neglecting neutrino shear stress. We compare them to exact numerical solutions computed by a Boltzmann code, and we find the two to be in very good agreement. The analytic solutions show explicitly that CDM and the baryon-photon fluid effectively behave as separate self-gravitating fluids until the epoch of baryon drag. This in turn leads to the surprising conclusion that the CMB is much less sensitive to the clustering properties of minimally coupled Dark Matter models than what would be naively expected.}
\maketitle

\section{Introduction and motivations}

Linear cosmological perturbation theory aims at describing the evolution of small deviations from the homogeneous Friedmann-Lema\^{\i}tre background up to times and scales when gravitational collapse leads to structure formation. This system of equations is essential in fitting high precision CMB data from e.g. Planck~\cite{Planck} as well as large scale structure data. The system cannot be solved straightforwardly, and precise calculations involve numerical Boltzmann codes. The derivation of analytic approximations is usually involved, but remains the best way to get some insight on the underlying physical mechanisms.

Cold dark matter is coupled to all other species through gravity. Its density fluctuation $\delta_c$ obeys
\be \ddot{\delta}_c + \frac{\dot{a}}{a} \dot{\delta}_c = 4 \pi G a^2 \sum_{\alpha} (\rho_\alpha + 3 P_\alpha ) \delta_\alpha~, \label{CDMevolution}\ee
where the sum is taken over all the species $\alpha$ \footnote{In equation (\ref{CDMevolution}), it is assumed that all species have adiabatic initial conditions.}. In 1974, P. M\'esz\'aros used this equation for describing the evolution of point-like masses in a homogeneous radiation background~\cite{Meszaros}. M\'esz\'aros simply neglected radiation perturbations thus making cold dark matter \emph{self-gravitating}, i.e.
\be \ddot{\delta}_c + \frac{\dot{a}}{a}\dot{\delta}_c - \frac{3}{2}\left(\frac{\dot{a}}{a}\right)^2 \Omega_c(a) \delta_c = 0~. \label{abc}\ee
This relation leads to two simple analytical solutions, which are \emph{a priori} valid in this framework only. However, the M\'esz\'aros equation has often been used to describe the evolution of cold dark matter in more general cases, when radiation perturbations actually dominate the perturbed stress-energy tensor and affect the gravitational potential. But as a matter of fact, the solution of the M\'esz\'aros equation matches numerical computation very well, even during radiation domination.

Thus, it appears that CDM is effectively gravitationally decoupled from other species.
It was not until 2002 that this result was supported by an analytical proof. By decomposing the perturbations into fast and slow modes, Weinberg~\cite{Weinberg2002} brought the first justification of why other contributions can be neglected, and proved that both the CDM and the photon-baryon fluid are essentially self-gravitating, even during radiation domination. The goal of this work is to better understand this gravitational decoupling using Weinberg's argument, and to verify the accuracy and the range of validity of the analytic approximations supporting this conclusion. 

This paper is organised as follows. In section 2, we present a complete, self-contained and purely analytical treatment of the small-scale perturbations. This part follows the approach of Weinberg. We work in a flat universe with no cosmological constant\footnote{Note that our solutions are equally valid for non-zero cosmological constant, since we are always restricting ourselves to times well before $\Lambda$-domination.}, neglecting the anisotropic shear of neutrinos and assuming tight-coupling between photons and baryons. In section 3, we compare our analytical solutions to the perturbations computed numerically by the Boltzmann code \CLASS{}~\cite{CLASS1,CLASS}. In section 4, we discuss to which extent these results prove that the CMB is insensitive to the clustering properties of Dark Matter. We illustrate the discussion with particular examples of Warm Dark Matter models, leaving no signature on the CMB in the observable range of angular scales, while these models actually suppress the growth of DM perturbations on scales that should in principle be observable. We summarise our results in section 5, and mention some possible applications of the analytic solutions for improving the efficiency of Boltzmann codes.

\section{Small-scale analytical solutions}
In this section, we solve the evolution of adiabatic scalar perturbations in the synchronous gauge until the time of recombination. We follow the approach proposed by Weinberg in 2002~\cite{Weinberg2002} and 2008~\cite{Weinberg2008}. In contrast with \cite{Weinberg2002} and like in \cite{Weinberg2008}, we consider photons and neutrinos separately. Furthermore, we derive new solutions for the slow modes of all species in the presence of baryons. 


For the sake of clarity, we adopt the same notations as Ma \& Bertschinger~\cite{MaBert}. They also correspond to the notations used by \CLASS{}. Throughout the rest of the paper, dots will denote derivatives with respect to conformal time $\tau$, and $\rho_\alpha$ ($P_\alpha$) will indicate the homogeneous background density (pressure) of the species $\alpha$. $\delta_\alpha \equiv \delta \rho_\alpha / \rho_\alpha$ will denote the relative density fluctuation, the velocity divergence is denoted by $\theta_\alpha$, and $R \equiv \frac{4 \rho_\gamma}{3 \rho_b}$ is the photon-to-baryon density ratio.

As long as we neglect neutrino shear and assume tight-coupling between photons and baryons ($\delta_b = \frac{3}{4} \delta_\gamma$ and $\theta_b = \theta_\gamma$), the three components (baryon+photon fluid, neutrinos and CDM) can be described by a set of continuity and Euler equations,
\begin{eqnarray}
\dot{\delta}_c = - \frac{\dot{h}}{2}~,
&& ~~~~~\dot{\delta}_\nu = - \frac{4}{3}\theta_\nu - \frac{2}{3}\dot{h}~,
~~~~~\dot{\delta}_\gamma = - \frac{4}{3}\theta_\gamma - \frac{2}{3}\dot{h}~, \label{Continuity}\\
\dot{\theta}_\nu = \frac{k^2}{4}\delta_\nu~,  
&&~~~~~(1+R)\dot{\theta}_\gamma +\frac{\dot{a}}{a}\theta_\gamma - k^2 R \frac{1}{4}\delta_\gamma = 0~. \label{Euler}
\end{eqnarray}
To close the system, we need an evolution equation for the trace $h$ of the metric perturbation,  inferred from Einstein's equation,
\begin{equation}
\ddot{h} + \frac{\dot{a}}{a} \dot{h} = - 8 \pi G a^2 \( \rho_c \delta_c + \( 2 + R^{-1}  \) \rho_\gamma \delta_\gamma + 2 \rho_\nu \delta_\nu \)~. \qquad \label{Einstein}
\end{equation}
Equations (\ref{Continuity}) -- (\ref{Einstein}) form a system of 6 equations with 6 unknowns. Unfortunately, it is not possible to find analytic solutions that would be valid for all times before decoupling, $\tau < \tau_\mr{dec}$, and all wavenumbers. However, asymptotic solutions can be found in two overlapping regions: the sub-Hubble region $k \tau \gg 1$, and the radiation domination era, when $\rho_R \gg \rho_M$ \footnote{ $\rho_R \equiv \rho_\gamma + \rho_\nu$ is the total energy density of relativistic species and $\rho_M \equiv \rho_b + \rho_D$ is the total density of non-relativistic species.}. For any wavenumber crossing the horizon during radiation domination ($k>k_{eq}$), a matching between the two regimes gives analytical solutions which are valid at any time before recombination.

We choose to normalise the scale factor at the time $\tau_{eq}$ of radiation-matter equality, $a(\tau_{eq}) = 1$. This is very convenient because $\rho_R = \rho_{eq}a^{-4}$ and $\rho_M = \rho_{eq} a^{-3}$, where $\rho_{eq} \equiv \rho_R(\tau_{eq}) = \rho_M(\tau_{eq})$. With this choice of normalisation\footnote{Comoving wavenumbers $k$ and conformal times $\tau$ are often reported in the literature in units such that $a(\tau_0)=1$. To compare with the quantities reported in this work, one should renormalise $k$ and $\tau$ by $a(\tau_{eq})/a(\tau_0)$.}, the radiation-domination condition $\rho_M \ll \rho_R $ is equivalent to $a \ll 1$.  By integrating the Friedmann equation, we can express the scale factor as a function of conformal time for a flat universe containing no cosmological constant, 
\be 
a(\tau) = \tau^2 \frac{2 \pi G \rho_{eq}}{3} + 2 \tau \sqrt{ \frac{2 \pi G \rho_{eq}}{3} }  \label{aoftau}.
\ee

\subsection{Radiation domination era}

Since $\rho_b$ and $\rho_\gamma$ have the same order of magnitude as $\rho_M$ and $\rho_R$, respectively, we have $R \gg 1$ during radiation domination. Therefore, the Euler equation for the photon-baryon plasma  reduces to the same equation as for neutrinos (see Eq.(\ref{Euler})). Thus,  $\delta_\gamma$ and $\delta_\nu$, as well as $\theta_\gamma$ and $\theta_\nu$, have the same equations of motion. Since they also share the same initial conditions, we have 
\begin{equation*}
\delta_\gamma = \delta_\nu \equiv \delta_R \qquad \mathrm{and} \qquad \theta_\gamma = \theta_\nu \equiv \theta_R. 
\end{equation*}
Note that $\rho_M \ll \rho_R$, but this does not imply that the dark matter perturbations are negligible in the Einstein equation (\ref{Einstein}). One might have $\delta_c \gg \delta_R$, which would counterbalance the fact that $\rho_M \ll \rho_R$, and prevent us from neglecting the matter source terms. For each wavenumber $k$, there is a time at which this becomes true (soon before equality for small $k$'s, or significantly before for large $k$'s). However, we will see \emph{a posteriori} that there always exists an epoch at which dark matter perturbations are still negligible, while at the same time, the solutions discussed in this section can be matched with the sub-Hubble solutions. Hence, we will assume for the time being that $\rho_M \delta_M \ll \rho_R \delta_R$. To summarise, the system of equations in the radiation dominated limit is the following:
\begin{eqnarray}
&&\dot{\delta}_c = - \frac{\dot{h}}{2}~,
~~~~~\dot{\delta}_R = - \frac{4}{3}\theta_R - \frac{2}{3}\dot{h}~,
~~~~~\dot{\theta}_R = \frac{k^2}{4}\delta_R~,   \label{RD4} \\
&& \ddot{h} + \frac{\dot{a}}{a} \dot{h} = - 16 \pi G a^2 \rho_R \delta_R \label{RD1}~.
\end{eqnarray}
Combining these equations and using (\ref{aoftau}) in the radiation-dominated limit $a\propto \tau$, we find a linear third order differential equation for $\dot{h}$:
\be \tau^2 \frac{d^3 \dot{h}}{d\tau^3} + 5 \tau  \frac{d^2 \dot{h}}{d\tau^2} + \(\frac{d\dot{h}}{d\tau}+\frac{\dot{h}}{\tau} \)\frac{k^2 \tau^2}{3}   = 0~. \ee
A basis of three linearly independent solutions for $\dot{h}$ is given by:
\be \dot{h}_1 \propto \frac{2 + \theta^2}{\theta^3},  \qquad \dot{h}_{2} \propto \frac{\sin{\theta} - \theta \cos \theta }{\theta^3}, \qquad  \dot{h}_{3} \propto \frac{\cos \theta + \theta \sin \theta}{\theta^3}, \ee
where $\theta \equiv \frac{k \tau}{\sqrt{3}}$. The general solution is given by $\dot{h}(\tau) = \mathcal{C}_1\dot{h}_1 + \mathcal{C}_2 \dot{h}_2 + \mathcal{C}_3 \dot{h}_3$, where the $\mathcal{C}_i$ are arbitrary real functions of $k$. We can identify the fastest growing solution in the small $\theta$ limit by considering the Taylor expansion of the general solution around $\theta=0$:
\be  \dot{h}(\tau) = (2 \mathcal{C}_1+\mathcal{C}_3) \( \frac{1}{\theta ^3}+ \frac{2}{\theta} \) + \frac{\mathcal{C}_2}{3} - \frac{\mathcal{C}_3}{8} \theta + \mathcal{O}(\theta^2), \ee
which shows that the choice $\mathcal{C}_3 = -2\mathcal{C}_1$ and $\mathcal{C}_2=0$ yields the fastest growing solution. The latter can be written as
\be \dot{h} = \frac{N}{\tau^3} \left( \cos \theta + \theta \sin \theta - 1 - \frac{\theta^2}{2} \right), \label{eq:rdh} \ee
where $N(k)$ is a normalisation factor. In the limit $\tau \longrightarrow 0$, all terms in $\theta^{-1}$ and $\theta^{-3}$ cancel out, and $\dot{h}\propto \theta \propto \tau$. One can show that this solution corresponds to an asymptotically constant comoving curvature perturbation in the super-Hubble limit: hence it is the usual growing adiabatic mode. We may plug the solution (\ref{eq:rdh}) and its time derivative into equations (\ref{RD4}, \ref{RD1}) to infer $\delta_R$, $\delta_c$ and $\theta_R$. The solutions are given in appendix~\ref{sec:ARD}. In the sub-Hubble limit, they reduce to
\begin{eqnarray}
\delta_R && \xrightarrow{\theta \to \infty} - \frac{Nk^2}{18} \cos \theta~, \label{eq:RD_R_limit}\\
\delta_c && \xrightarrow{\theta \to \infty}      \frac{Nk^2}{12} \left(  \gamma +  \log( \theta ) - \frac{1}{2} \right) ~, \label{eq:RD_C_limit}
\end{eqnarray}
where $\gamma = 0.57721 \dots$ is the Euler-Mascheroni constant. Then, the ratio $R_{CR}$ between the absolute density fluctuations of CDM and of radiation behaves inside the Hubble radius like
\begin{equation}
R_{CR} \equiv \frac{\rho_c \delta_c}{\rho_R \delta_R} \propto \tau^2 \log(k \tau)~.
\end{equation}
Therefore, the amount of time between Hubble crossing ($\tau \sim 1/k$) and the take-over of CDM fluctuations ($R_{CR}=1$) increases with $k$. Hence, for growing $k$, there is an increasing range of time available for doing the matching between the radiation-dominated asymptotic solution (following from $a \ll 1$, $R_{CR} \ll 1$) and the sub-Hubble asymptotic solution (following from $\theta \gg 1$). This confirms \emph{a posteriori} the validity of the approach that we are following.

\subsection{Deep inside the Hubble radius}

We now consider modes which are well inside the Hubble scale, $k \tau \gg 1$. 
The system (\ref{Continuity}-\ref{Einstein}) can be solved approximately by decomposing the solutions into fast and slow modes, defined as modes which evolve at a time-scale of $k$ and $\tau^{-1}$ respectively:
\be A = A^{\mathrm{fast}} + A^{\mathrm{slow}} \qquad \mathrm{with} \qquad \frac{d\log A^{\mathrm{fast}} }{d\tau}  = \mathcal{O} \left(k \right), \qquad  \frac{d \log A^{\mathrm{slow}} }{d\tau}  = \mathcal{O} \left(\frac{1}{\tau} \right).\nonumber\ee
Since we consider three fluids, we expect six independent physical solutions. We will see that the decomposition in fast and slow modes leads precisely to this number of solutions. 
We will find that photon and baryon densities are dominated by fast modes, by two powers of $\theta$. In contrast, cold dark matter is strongly dominated by slow modes, also by two powers of $\theta$.

\paragraph{Fast modes}
\paragraph{}
\noindent Replacing conformal time derivatives by $k$, we see that the left-hand side of the Einstein equation (\ref{Einstein}) is of order $k^2 h^{\mr{fast}}$. In contrast, the Friedmann equation implies that the first term of the right-hand side is at most of order $\frac{\dot{a}^2}{a^2} \delta_c^{\mr{fast}} \sim \tau^{-2} \delta_c^{\mr{fast}}$. But the continuity equation for CDM shows that $h$ and $\delta_c$ are of the same order of magnitude. Therefore, the ratio of these two quantities is of order $\theta^{2} \gg 1$, and we can safely neglect $\delta_c^{\mr{fast}}$ in  equation (\ref{Einstein}).
Moreover, the neutrino contribution $\delta_\nu^{\mr{fast}}$ to the  gravitational potential can also be neglected because of their non-zero anisotropic shear, which carries the amplitude of perturbations towards higher multipole moments in the limit $k \tau \gg 1$. 

Consequently, regarding the fast modes, the source term of the Einstein equation (\ref{Einstein}) is completely dominated by the photon-baryon plasma. It follows that $\delta_\gamma^{\mr{fast}}$ is at least of order $k^2 \tau^2 h^{\mr{fast}}$. From the Euler equation, we see that $\theta_\gamma^{\mr{fast}}$ is of order $k \delta_\gamma^{\mr{fast}}$. Thus, in the continuity equation, the left-hand side is of order $k \delta_\gamma^{\mr{fast}} \ge (k \tau)^2 kh^{\mr{fast}} \gg k h^{\mr{fast}}$, the first term on the right is also of order $\theta_\gamma^{\mr{fast}} \approx k \delta_\gamma^{\mr{fast}} \gg k h^{\mr{fast}}$, whereas the second term on the right is of order $kh^{\mr{fast}}$. Hence we can safely neglect the metric source term in the photon continuity equation.
In conclusion, the system governing the evolution of fast modes reads
\begin{eqnarray}
&&  \dot{\delta}_c^{\mr{fast}} = - \frac{\dot{h}^{\mr{fast}}}{2}~,
~~~~~\dot{\delta}_\nu^{\mr{fast}} = - \frac{4}{3}\theta_\nu^{\mr{fast}} - \frac{2}{3}\dot{h}^{\mr{fast}}~,
~~~~~\dot{\delta}_\gamma^{\mr{fast}} = - \frac{4}{3}\theta_\gamma^{\mr{fast}}~,  \label{f_continuity}   \\
&&  \dot{\theta}_\nu^{\mr{fast}} =  \frac{k^2}{4}\delta_\nu^{\mr{fast}}~,
~~~~~(1+R)\dot{\theta}_\gamma^{\mr{fast}} + \frac{\dot{a}}{a}\theta_\gamma^{\mr{fast}} - k^2 R \frac{1}{4}\delta_\gamma^{\mr{fast}} = 0 \label{f_euler}~, \\
&&  \ddot{h}^{\mr{fast}} + \frac{\dot{a}}{a} \dot{h}^{\mr{fast}} = - 8 \pi G a^2 \rho_\gamma(2 + R^{-1}) \delta_\gamma^{\mr{fast}}~.   \label{f_einstein}
\end{eqnarray}
%
We can show that this system has four independent solutions. Combining the Euler and continuity equation, we obtain a relation for $\delta_\gamma$ alone:
\be  (1+R)\ddot{\delta}_\gamma^{\mr{fast}} + \frac{\dot{a}}{a} \dot{\delta}_\gamma^{\mr{fast}} + \frac{1}{3}k^2 R \delta_\gamma^{\mr{fast}} = 0~. \label{WaveEq} \ee
This equation has two independent solutions, which fully determine $\theta_\gamma^{\mr{fast}}$. Since the homogeneous solution of equation~(\ref{f_einstein}) is a slow mode, $\dot{h}^{\mr{fast}}$ is also fully determined by  $\delta_\gamma^{\mr{fast}}$. Similarily, since a constant $\delta_c$ would clearly not be a fast mode, ${\delta}_c^{\mr{fast}}$ is also fully determined by  $\delta_\gamma^{\mr{fast}}$. Finally, solving for $\delta_\nu^{\mr{fast}}$ and $\theta_\nu^{\mr{fast}}$ leads to another second-order differential equation bringing two more independent solutions. 

One of the noteworthy aspects of these equations is that the fast mode of the photon perturbations is completely independent of the dark matter perturbations, whereas equation (\ref{WaveEq}) applies at any time, even during the matter-dominated era. Since $a$ and $R$ are by definition slowly varying, we can solve Eq.~(\ref{WaveEq}) with the WKB approximation
\be \delta_\gamma^{\mr{fast},\pm} = \frac{1}{(1+R^{-1})^{\frac{1}{4}}} \exp \left(\pm i k r_s\right), \ee
where $r_s$ is the sound horizon, 
\be
r_s = \int^\tau  c_s \, d\tilde{\tau} = \int^\tau \left(\frac{R}{3(1+R)}\right)^{1/2} d\tilde{\tau}~,
\ee
which can be solved analytically in a $\Lambda = 0$ universe, 
\be r_s =  \int_0^\tau \sqrt{\frac{R}{3(1+R)}} d\tilde{\tau}  = \frac{1}{\sqrt{8 \pi G \rho_{eq}}} \frac{2}{\sqrt{K}}  \log \left( \frac{K\sqrt{1+a} + \sqrt{K(1+Ka)}}{K + \sqrt{K}} \right) ,    \ee
where $K$ is the constant such that $R^{-1}=Ka$.  These solutions describe the well-known photon acoustic oscillations, damped by baryon inertia when $R$ is of order one.

Any linear combination of $\delta_\gamma^{\mr{fast}, +}$ and $\delta_\gamma^{\mr{fast}, -}$ is also a solution to equation (\ref{WaveEq}). For wavenumbers $k > k_{eq}$, we can pick the unique linear combination by matching with the adiabatic growing mode solution in the radiation-dominated limit (equation (\ref{sol:GammaRD}). This provides the general solution for fast modes crossing the Hubble scale before radiation-matter equality. Photon density fluctuations are given by
\be \delta_\gamma^{\mr{fast}} = - \frac{Nk^2}{18(1+R^{-1})^\frac{1}{4} } \cos(kr_s)~, \label{DeltaGammaFast}\ee
and uniquely determine $\dot{h}^{\mr{fast}}$, ${\delta}_c^{\mr{fast}}$ and $\theta_\gamma^{\mr{fast}}$, as shown in Appendix~\ref{sec:Match_slow}. The solution for strongly damped neutrino fast modes could be obtained in a similar way.

These results are valid inside the Hubble radius, as long as baryons and photons are tightly coupled. To properly account for the evolution of small wavelengths, one must also take Silk damping into account, i.e. the fact that the photon mean free path is finite and increases with time.
This results in an exponential decay of $\delta_\gamma^{\mr{fast}}$, which to a good approximation is captured by the following factor~\cite{Weinberg2008}:
%
\be \delta_\gamma^{\mr{fast}} \to \delta_\gamma^{\mr{fast}} \times \exp \left( -\int_0^\tau \frac{k^2 }{6 a (1+R^{-1}) \sigma_T n_e } \left( \frac{16}{15} + \frac{1}{R(R+1)} \right) d\tilde{\tau} \right),\label{silk}\nonumber\ee
where $n_e(a)$ is the free electron density.

\paragraph{Slow modes}
\paragraph{}
\noindent We will now carry out a similar analysis for slow modes. The Euler equation for neutrinos (\ref{Euler}) shows that $\delta_\nu^{\mr{slow}}$ is of order $\frac{\theta_\nu^{\mr{slow}}}{k^2 \tau}$. Thus, in the continuity equation (\ref{Continuity}), the left-hand side is of order $\dot{\delta}_\nu^{\mr{slow}} \approx \frac{\theta_\nu^{\mr{slow}}}{k^2 \tau^2} \ll \theta_\nu^{\mr{slow}}$ and can be dropped. Similarly, for the photon-baryon plasma,  the left-hand side of the continuity equation (\ref{Continuity}) is much smaller than $\theta_\gamma^{\mr{slow}}$ and can be dropped as well.

Consequently,  $\theta_\nu^{\mr{slow}}$ and $\theta_\gamma^{\mr{slow}}$ are both of order $h^{\mr{slow}}/\tau$. Then the Euler equations (\ref{Euler}) imply that $\delta_\nu^{\mr{slow}}$ and $\delta_\gamma^{\mr{slow}}$ are of order $\frac{h^{\mr{slow}}}{k^2 \tau^2}$ and $\frac{h^{\mr{slow}}}{k^2 \tau^2}\frac{2+R}{R}$, respectively. In contrast, the dark matter perturbation is of order $\delta_c^{\mr{slow}} \approx h^{\mr{slow}}$ according to its continuity equation (\ref{Continuity}). Therefore, the Einstein equation (\ref{Einstein}) is strongly dominated by $\delta_c^{\mr{slow}}$  and all other contributions can be dropped. We obtain the following system of equations for slow modes:
\begin{eqnarray}
&&  \dot{\delta}_c^{\mr{slow}} = \theta_\nu^{\mr{slow}}  = \theta_\gamma^{\mr{slow}} = - \frac{\dot{h}^{\mr{slow}}}{2}~,\label{slow_continuity} \\
&& \dot{\theta}_\nu^{\mr{slow}} =  \frac{k^2}{4}\delta_\nu^{\mr{slow}}~,
\qquad (1+R)\dot{\theta}_\gamma^{\mr{slow}} + \frac{\dot{a}}{a}\theta_\gamma^{\mr{slow}} - k^2 R \frac{1}{4}\delta_\gamma^{\mr{slow}} = 0~, \label{slow_euler} \\
&& \ddot{h}^{\mr{slow}} + \frac{\dot{a}}{a}\dot{h}^{\mr{slow}} = - 8 \pi G a^2 \rho_c \delta_c^{\mr{slow}}~. \label{slow_einstein}
\end{eqnarray}
Unlike for fast modes, the solutions are fully determined by $\delta_c^{\mr{slow}}$: knowing its evolution, one can infer $h^{\mr{slow}}$ and $\theta_{\gamma, \nu}^{\mr{slow}}$ from the continuity equations, and $\delta_{\gamma, \nu}^{\mr{slow}}$ from the Euler equations. To find $\delta_c^{\mr{slow}}$, we inserting the CDM continuity equation in (\ref{slow_einstein}), and obtain the second-order differential equation
\be  \ddot{\delta}_c^{\mr{slow}} + \frac{\dot{a}}{a} \dot{\delta}_c^{\mr{slow}} = 4 \pi G a^2 \rho_c \delta_c^{\mr{slow}} ~,
\label{eqCDM}\ee
which has two independent solutions as usual. Together with the four independent fast solutions, they complete the full set of six solutions. 

Equation (\ref{eqCDM}) describes the clustering of cold dark matter before the baryon drag epoch (i.e. as long as baryons are tightly coupled to photons). The right-hand side corresponds to the gravitational force that attracts CDM into over-dense regions. The second term of the left-hand side is the Hubble friction, which accounts for the fact that expansion slows down the clustering process. 

\begin{figure}[ht!]
\centerline{\includegraphics[scale=0.6]{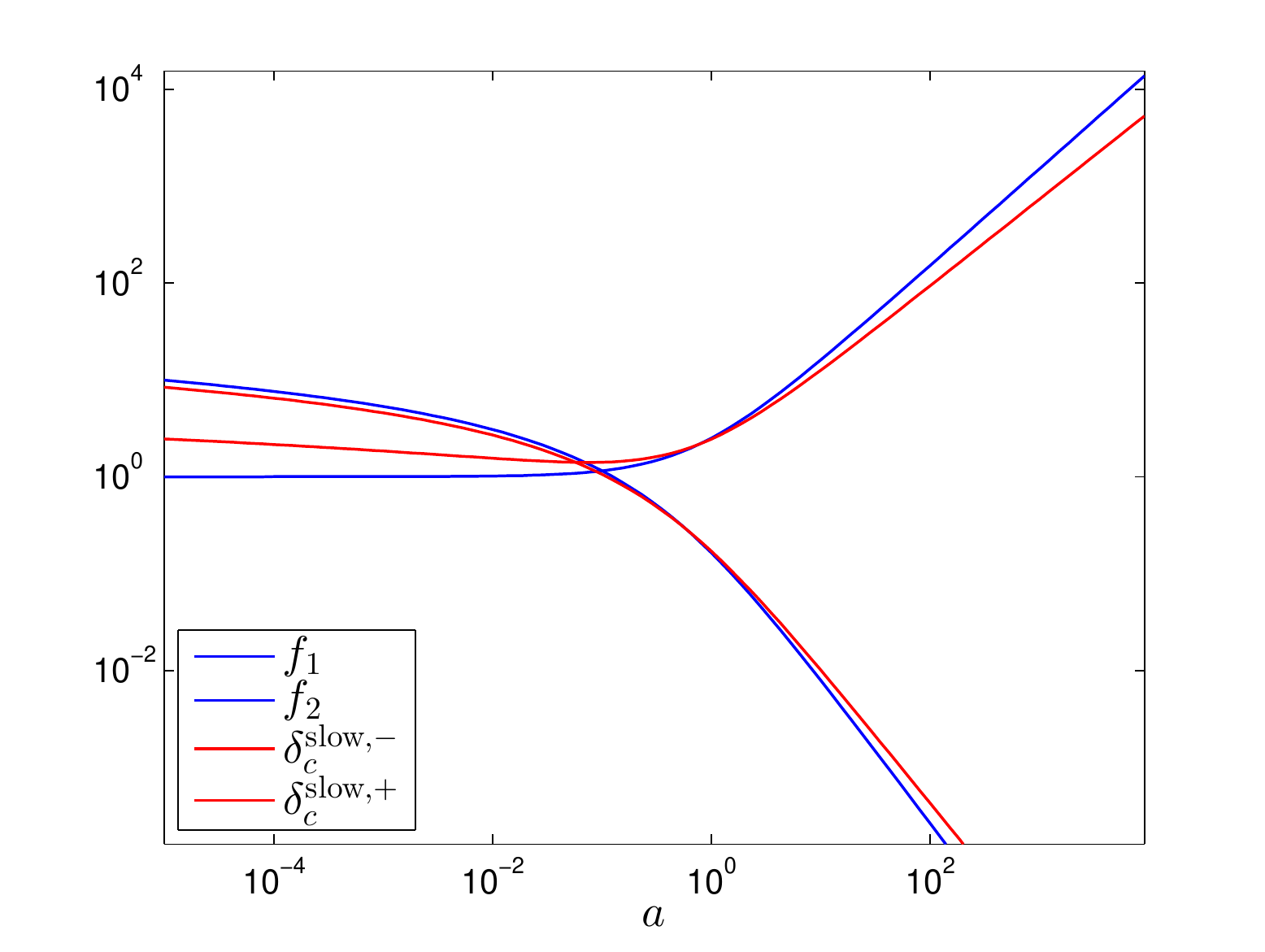}}
\caption{Comparison between the exact and simplified solutions for $\beta = 1/6$.\label{fig:bla} }
\end{figure}

The crucial point is that the gravitational force acting on CDM \emph{does not contain any contribution from the baryon-photon plasma}, although equation (\ref{eqCDM}) applies on sub-Hubble scales at any epoch, \emph{even during radiation domination}. In fact, this relation correspond to the self-gravitating M\'esz\'aros equation~(\ref{abc}). But it is important to note that equation~\eqref{eqCDM} has been derived without the assumption of negligible radiation perturbations which led to the M\'esz\'aros equation\footnote{We stress that it is only the \emph{slow modes} of CDM which are gravitationally decoupled from other species. But as we will see in section \ref{sec:summary}, fast modes are completely negligible for CDM.}. Using the Friedmann equation and introducing the baryon-to-matter density ratio $\beta \equiv \Omega_b / \Omega_M$, we can write equation~(\ref{eqCDM}) in terms of $a$:  
\be 
a(a+1)\frac{d^2 \delta_c^{\mr{slow}}}{da^2} + \left( 1 + \frac{3 a}{2} \right) \frac{d \delta_c^{\mr{slow}}}{da} - \frac{3}{2}(1-\beta)\delta_c^{\mr{slow}} = 0~. 
\ee
The independent solutions for $\beta = 0$ were first given by M\'esz\'aros in 1974 \cite{Meszaros}, and Groth \& Peebles in 1975
\cite{GP75}:
\be 
f_1(a) =  1 + \frac{3a}{2}~, \qquad f_2(a) = \left(1 + \frac{3a}{2}\right) \log \left( \frac{\sqrt{1+a} +1}{\sqrt{1+a}-1} \right) - 3 \sqrt{1+a} ~. \nonumber
\ee
In the radiation-domination limit $a \longrightarrow 0$, $f_1$ tends to $1$, whereas $f_2$ tends to infinity. On the contrary, when $a \longrightarrow \infty$, $f_1$ tends to infinity while $f_2$ is quickly suppressed. The solutions for any constant $\beta$ can be written in terms of hypergeometric functions (see \cite{HuSu96}): 
\be 
\delta_c^{\mr{slow},\pm}(a) \propto (1+a)^{-\alpha_{\pm}} \,_2F_1\left(\alpha_{\pm},\alpha_{\pm}+\frac{1}{2};2 \alpha_{\pm} + \frac{1}{2};\frac{1}{1+a}\right), \label{eq:smallscalecdm}
\ee 
where $\,_2F_1(a,b;c;z)$ is the Gauss hypergeometric function and 
\be
\alpha_{\pm} \equiv \frac{1 \pm \sqrt{1+24(1-\beta)}}{4} ~.
\ee
The proportionality factors are $3/2$ for $\delta_c^{\mr{slow},-}$ and $4/15$ for $\delta_c^{\mr{slow},+}$. If $\beta$ is small, we have at first order $\delta_c^{\mr{slow},-} \propto a^{1-\frac{3}{5} \beta}$ and $\delta_c^{\mr{slow},+} \propto a^{-\frac{3}{2}(1-\frac{2}{5} \beta)}$. Consequently, the presence of tightly coupled baryons slows down the growth of dark matter perturbations. The exact ($\beta\neq0$) and simplified ($\beta = 0$) solutions are compared in figure \ref{fig:bla} for $\beta = 1/6$. It appears that neglecting the effect of baryons is a very rough approximation, particularly in the matter dominated epoch. 

The M\'esz\'aros equation is a second-order linear differential equation, leading to two linearly independent solutions. For wavenumbers $k > k_{eq}$, we can use the radiation-domination result (\ref{eq:RD_C_limit}) to determine the unique solution corresponding to the growing adiabatic mode. We match the radiation dominated limit of the small scale solution~\eqref{eq:smallscalecdm} to the sub-Hubble limit of the radiation domination solution, equation~\eqref{eq:RD_C_limit}. The solution (\ref{eq:RD_C_limit}) is a slow mode and the matching condition reads
\be \lim_{\theta \to \infty} \delta_c^{\mr{RD}} = \lim_{a \to 0} ( \mathcal{A} \delta_c^{\mr{slow},+} + \mathcal{B} \delta_c^{\mr{slow},-} )~. 
\label{eq:matching}
\ee
We find
\be
\delta_c^{\mr{slow}} = \left\{ 
\begin{array}{ccc}
\frac{Nk^2}{12}  \left[ \left(  \gamma + \log\left(\frac{2q}{\sqrt{2\pi G \rho_{eq}}}\right) -\frac{7}{2}\right)f_1 - f_2 \right] \qquad &\mr{for}&\qquad \beta = 0  \\[24pt]
\frac{2}{3} \mathcal{C}(\alpha_{+}, \alpha_{-} )  \delta_c^{\mr{slow},-} + \frac{15}{4} \mathcal{C}(\alpha_{-}, \alpha_{+} ) \delta_c^{\mr{slow},+} \qquad&\mr{for}&\qquad \beta \neq 0
\end{array}
\right.
\label{DeltaCSlow}
\ee
where $\mathcal{C}$ is a function of $k$ involving Euler gamma and di-gamma functions. The derivation of this result, the definition of the function $\mathcal{C}$ and the solution for other quantities are all presented in Appendix~\ref{sec:Match_slow}.

\subsection{Summary\label{sec:summary}} 
We have found a set of small-scale solutions, each of them containing a slow and a fast contribution. We can now compare them, and find which are the dominant processes. Comparing equations (\ref{app:FinalDeltaCFast}) with (\ref{app:FinalDeltaCSlow_simplified}) and (\ref{app:FinalDeltaCSlow}), one can show that $\delta_c^\mr{slow}$ is bigger than $\delta_c^{\mr{fast}}$, at least by a factor $\theta^{2}$. Thus, cold dark matter is strongly dominated by slow modes: $\delta_c \approx \delta_c^\mr{slow}$. Its evolution is entirely described by the the M\'esz\'aros equation (\ref{eqCDM}), in which the source term does not contain any contribution from other species. Consequently, cold dark matter is effectively self-gravitating, even during radiation domination, for times and scales such that $R_{CR} = (\rho_c \delta_c)/(\rho_R \delta_R) \ll1$. The reciprocal is also true for the baryon-photon plasma: $\delta_\gamma$ is dominated by the fast modes\footnote{This is also true for the velocity perturbation, but the ratio is only of one power of $\theta$.}, also by two orders of magnitude of $\theta$. Thus, photon perturbations are governed by the wave equation (\ref{WaveEq}), and do not experience any significant gravitational interaction with cold dark matter before decoupling, even when $R_{CR}  \gg1$.

Therefore, we conclude that cold dark matter and the photon-baryon plasma are effectively gravitationally decoupled. From a mathematical point of view, this stems from the decomposition into fast and slow modes. Physically, this comes from the fact that $\delta_\gamma$ oscillates quickly around zero, as shown by equation (\ref{DeltaGammaFast}), while cold dark matter is slowly collapsing. During the characteristic evolution time of CDM (the Hubble time $\tau^{-1}$), $\delta_\gamma$ undergoes a lot of oscillations with zero average value. The plasma wave fronts travel so fast that dark matter does not have time to feel its gravitational impact. The reciprocal effect comes from the fact that photon experience pressure forces that are much stronger than gravitational forces. The slow collapse of CDM could in principle shift the zero-point of photon oscillations away from zero, but in practise this shifting is suppressed by two powers of $\theta$ with respect to the amplitude of the oscillations. Hence it remains completely negligible on sub-Hubble scales.


\section{Numerical simulation of the perturbations}
We are now going to compare our analytical results (equations (\ref{DeltaGammaFast}),(\ref{DeltaCSlow}), (\ref{app:FinalHDotFast}) -- (\ref{app:FinalThetaGammaFast}) and (\ref{app:FinalOtherSlowModes_simplified}) -- (\ref{app:FinalDeltaGammaSlow})) with the exact numerical solutions computed using the Boltzmann code \CLASS{}~\cite{CLASS1,CLASS}. This will allow us to explicitly test their accuracy and to confirm our interpretation of the gravitational decoupling. The photon-to-baryon ratio $R$ is calculated with the usual thermodynamical relation. The normalisation factor $N$ is inferred from the initial condition used by \CLASS{} for adiabatic modes. We have $ \delta_\gamma (\tau_\text{ini}) = \frac{1}{3} \tau^2 k^2 $, so from the limit $\theta \to 0$ of equation (\ref{sol:GammaRD}) we identify $ N = 72 /k^2$.

In all figures of this section, we will show the evolution of the wavenumber $k=10.6 h/\text{Mpc}$. This choice has been made mainly for illustrative purposes. Indeed, by picking up such a small wavelength, one can see very well the different stages of evolution of perturbations, and visualise many acoustic oscillations in the baryon-photon fluid. Still, the solutions derived in section~2 are close to numerical results also on larger wavelengths.

\subsection{Cold dark matter}

\begin{figure}[h!]
\begin{minipage}{0.5\linewidth}
\centerline{\includegraphics[scale=0.4]{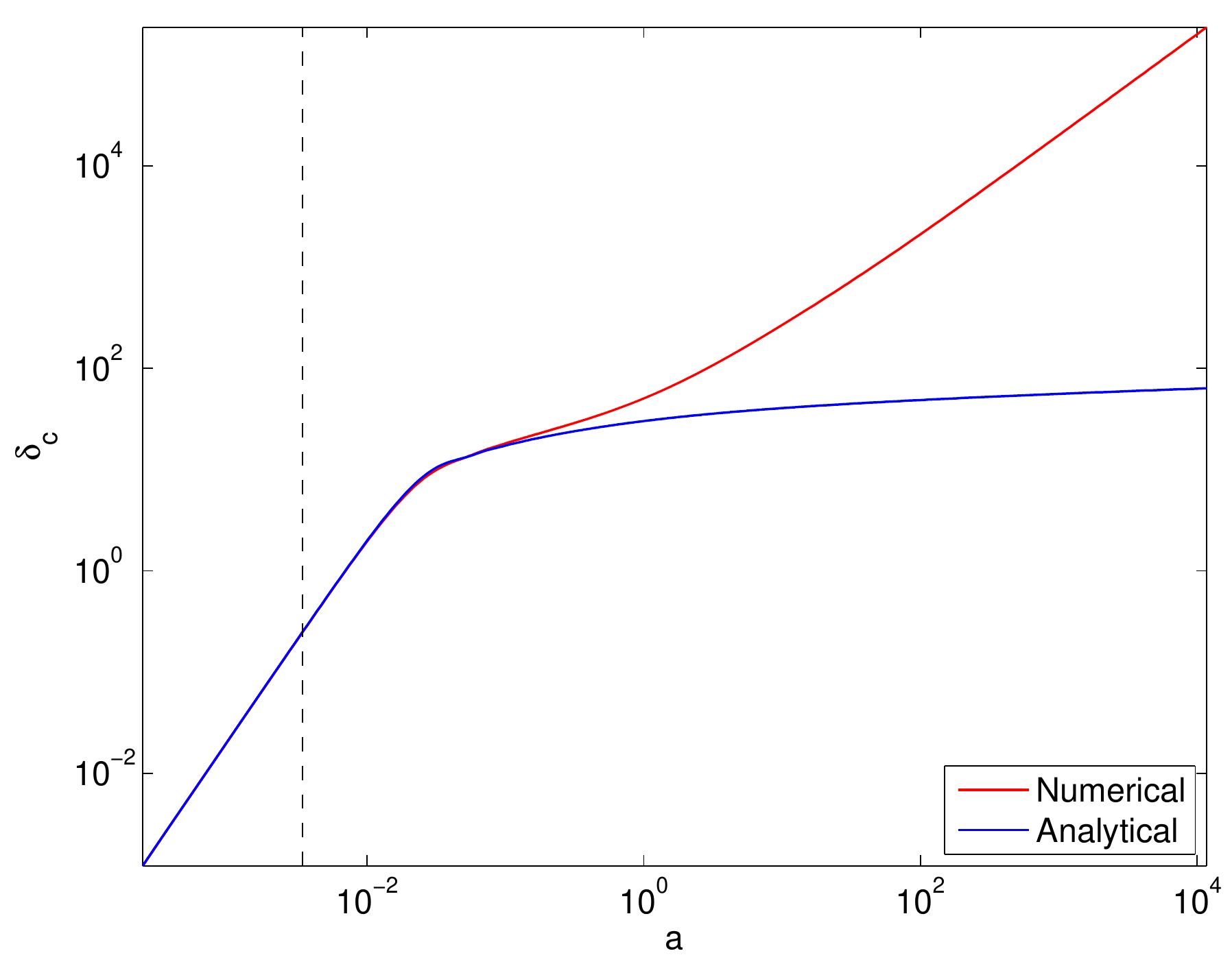}}
\end{minipage}
\hspace{0.15cm}
\begin{minipage}{0.5\linewidth}
\centerline{\includegraphics[scale=0.4]{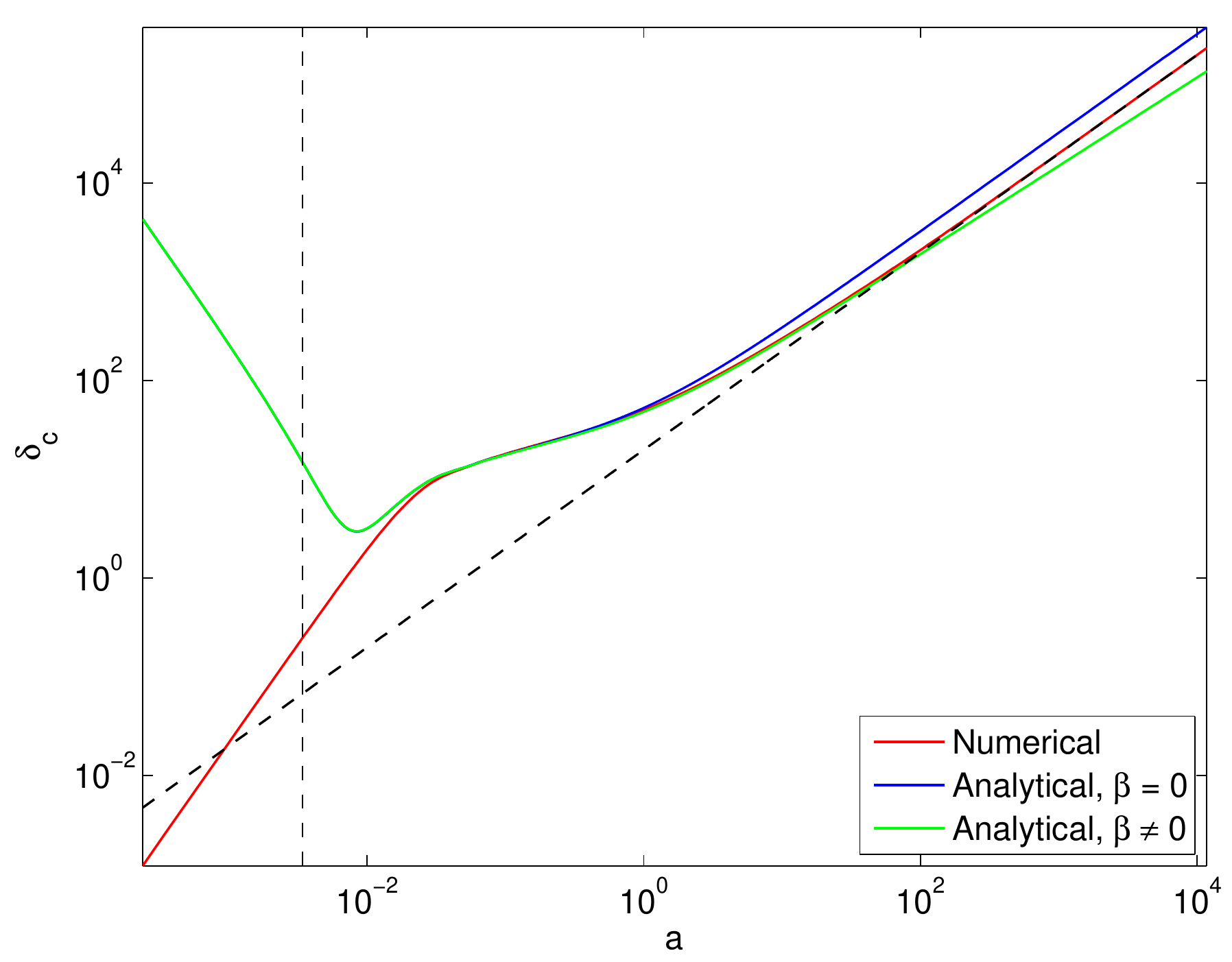}}
\end{minipage}
\caption{The radiation-domination \emph{(left panel)} and sub-Hubble \emph{(right panel)} solutions for cold dark matter are compared with numerical simulations. We have taken a wavenumber $k=10.6 h/\text{Mpc}$ and $\beta = 1/6$. The vertical dashed line corresponds to Hubble crossing. The diagonal dashed line is proportional to $a$. }
\label{fig:cdm}
\end{figure}

We will first consider cold dark matter. The radiation-domination solution~(\ref{sol:CDMRD}) and sub-Hubble solutions~(\ref{DeltaCSlow}) are compared with numerical simulations in figure~\ref{fig:cdm}.

The radiation-domination solution in the left panel of figure~\ref{fig:cdm} closely follows the numerical solution for $a \lesssim 0.1$, but then it diverges when matter starts to become important. On the right panel we show the $\beta = 0$ and $\beta \neq 0$ sub-Hubble solutions. They behave asymptotically as $\delta_c \propto a $ and $\delta_c \propto a^{1-\frac{3}{5}\beta}$ respectively. Since they reach the numerical solution before the radiation-domination solution becomes inexact, we are able to describe the density perturbation for all values of $a$.

The exact and simplified sub-Hubble solutions match perfectly in the radiation dominated era. This is surprising, because the approximate solution neglects the baryonic part of the \emph{background} matter, which is a rough simplification. Indeed, we can see in figure \ref{fig:bla} that the impact of $\beta \neq 0$ is not negligible. Therefore, we would expect the $\beta = 0$ solution to fit worse than the hypergeometric expression. In fact, during radiation domination, the source term of the M\'esz\'aros equation (\ref{eqCDM}) is sub-dominant. Indeed, as we are considering slow modes, the two terms of the left-hand side are of order $\tau^{-2} \delta_c^{\mr{slow}}$. In contrast, the right-hand side is of order $\tau^{-2} a \delta_c^{\mr{slow}}$. Thus, in the limit $a \to 0$, the evolution of cold dark matter is dominated by the rapid expansion of the universe, explaining why the two solutions are equivalent. 

\begin{figure}[h!]
\centerline{\includegraphics[scale=0.5]{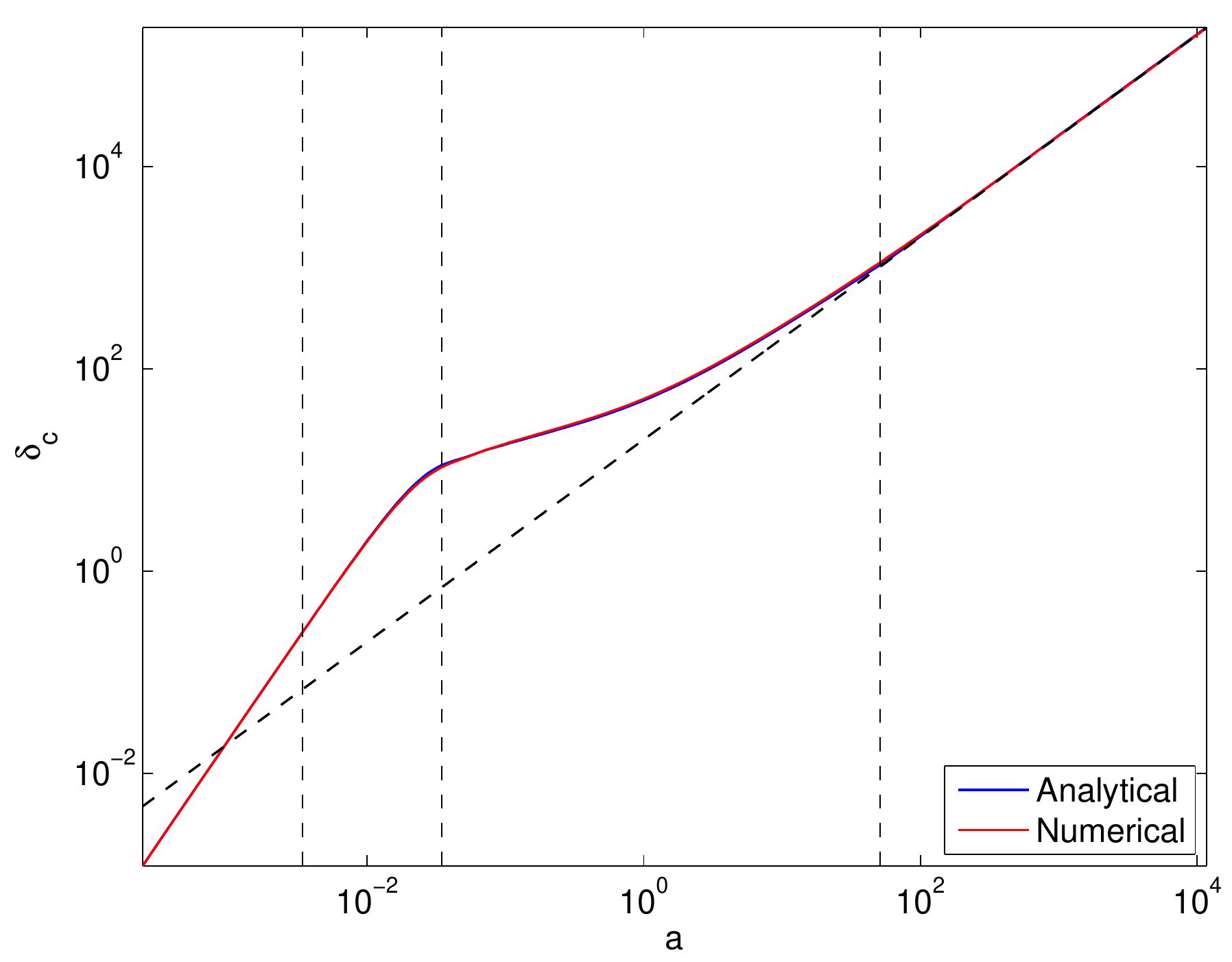}}
\caption{Concatenation of the three analytical solutions (radiation-domination, exact ($\beta \neq 0$) and simplified ($\beta =0$) ) for $k=10.6 h/\text{Mpc}$ and $\beta = 1/6$.}
\label{awgr}
\end{figure}

When $a\gtrsim  1$, we begin to see an effect from neglecting the background baryons: the simple analytical solution deviates from the numerical result. On the contrary, the hypergeometric solution matches the numerical simulation very well. This confirms the gravitational decoupling between the photon-baryon plasma and cold dark matter. For $a \gg 1$, we are outside the range of validity of our analytic solution (baryons and photons are no longer tightly coupled) and the hypergeometric solution no longer tracks the numerical solution. Indeed, after the drag time, baryons are no longer prevented by the photons from collapsing and fall freely into the gravitational potential wells. At that time, they behave as cold dark matter and $\delta_b^{\mr{slow}}$ increases. The perturbations of the two species equilibrate until $\delta_c^{\mr{slow}} = \delta_b^{\mr{slow}} = \delta_M$. Then, the Einstein and continuity equation yield\footnote{This equation could also have been obtained by setting $\beta \to 0$, as baryons behave like cold dark matter.}
\be \ddot{\delta}_M + 2 \frac{\dot{a}}{a} \dot{\delta}_M =   4\pi G \rho_{tot} \delta_M = \frac{3}{2}\left(\frac{\dot{a}}{a}\right)^2 \Omega_M(a) \delta_M. \ee
Thus, well after baryon drag, the evolution of CDM is ruled by the $\beta = 0$ solution. As a result, the numerical curve is parallel to the standard Meszaros solution and grows as $a$. To summarise, the hypergeometric solution describes the evolution of cold dark matter as long as the slow modes of baryons are negligible. As soon as $\delta_b^{\mr{slow}}$ becomes important, we have to use the $\beta = 0$ solution. In figure \ref{awgr}, we concatenate these two analytical solutions at a scale factor five times bigger than the drag time, while showing  the radiation-domination solution for $a \lesssim 0.1$. This allows us to describe analytically the evolution of cold dark matter at any time, inside and outside the Hubble radius.

\subsection{Photons and baryons}
\begin{figure}[h!]
\begin{minipage}{1\columnwidth}
\centerline{\includegraphics[width=\linewidth]{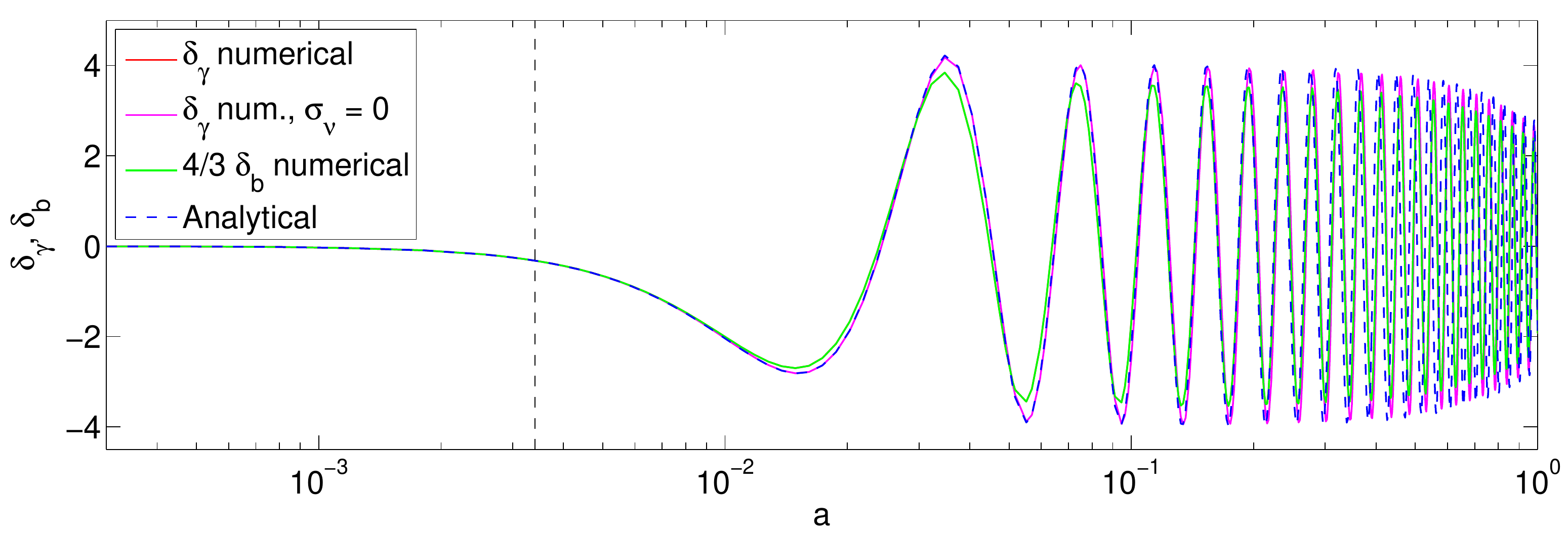}}
\end{minipage}
\begin{minipage}{1\columnwidth}
\centerline{\includegraphics[width=\linewidth]{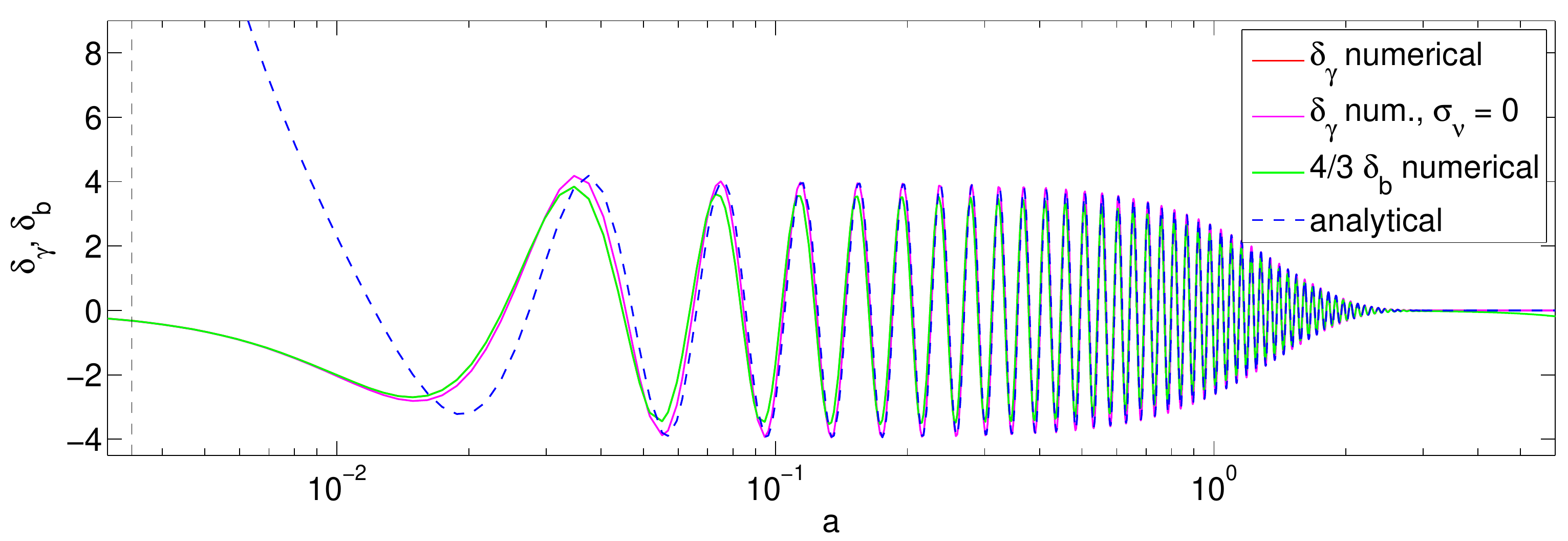}}
\end{minipage}
\caption{The radiation-domination (top panel) and sub-Hubble (bottom panel) solutions for the photon and baryon perturbations are compared with numerical simulations. We have taken a wavenumber $k=10.6 h/\text{Mpc}$ and calculated the Silk damping numerically. The slow mode is inferred from the exact hypergeometric solution.}
\label{fig:deltagamma}
\end{figure}

\begin{figure}[h!]
\begin{minipage}{1\columnwidth}
\centerline{\includegraphics[width=\linewidth]{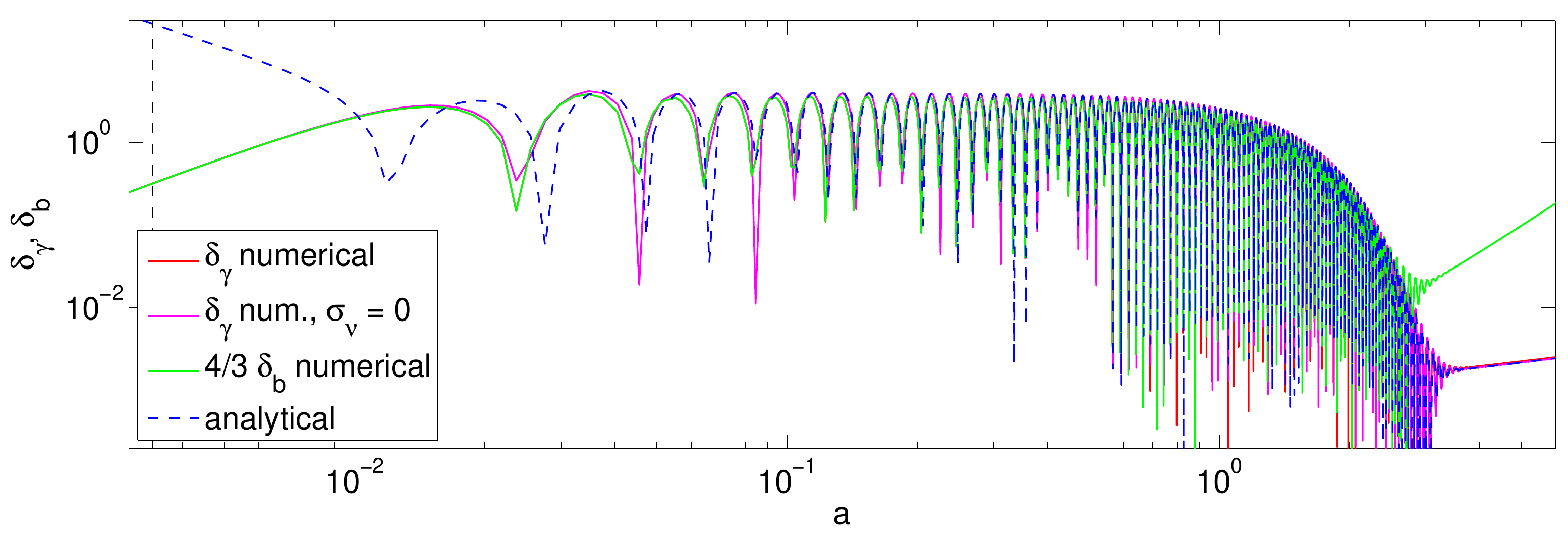}}
\end{minipage}
\caption{Sub-Hubble solutions for the photon and baryon perturbations, identical to the bottom panel of figure 4, but shown in logarithmic scale in order to see the emergence of slow modes after the washing out of fast modes by Silk damping.}
\label{fig:deltagammalog}
\end{figure}

The density perturbations of photons and baryons are shown in figure~\ref{fig:deltagamma} and~\ref{fig:deltagammalog} in linear and logarithmic scale respectively.  

When modes cross the Hubble radius, the density perturbations of the plasma start to oscillate around zero. Since we are considering very small wavelengths, they are soon damped by diffusion damping. After the drag time, the baryons start to collapse into the gravitational potential wells and their density perturbation increases. 

In the analytic solution, the Silk damping was computed using a numerical integral based on the free electron fraction computed by \CLASS{}. We calculated the numerical solution with \CLASS{} in two ways: without making approximations, or forcing the neutrino shear $\sigma_\nu$ to vanish, in order to match one of the simplifying assumptions done in the analytical calculation.

The analytical solution matches the numerical simulation with $\sigma_\nu=0$ very well once inside the Hubble radius. In particular, the amplitude and phase of the oscillations is remarkably well described. When comparing with the exact numerical solution with $\sigma_\nu \neq 0$, we see that the phase is still reproduced very well, but the amplitude is reduced by approximately $10\%$. The explanation is that the gravitational interaction between photons and neutrinos is important during a short range of time, soon after Hubble crossing~\cite{Lesgourgues:1519137}. During that time, any error on the neutrino solution (like the fact of treating them as a relativistic fluid instead of a free-streaming component experiencing damped oscillations) will get imprinted on the amplitude of photon oscillations.

The fact that odd and even maxima have the same amplitude in all solutions confirms that the slow modes are completely negligible. We stress that dark matter is completely absent in the analytical expression, but it still matches the numerical solution very well during matter domination.

\subsection{Velocity perturbation}

\begin{figure}[h!]
\begin{minipage}{1\columnwidth}
\centerline{\includegraphics[width=\linewidth]{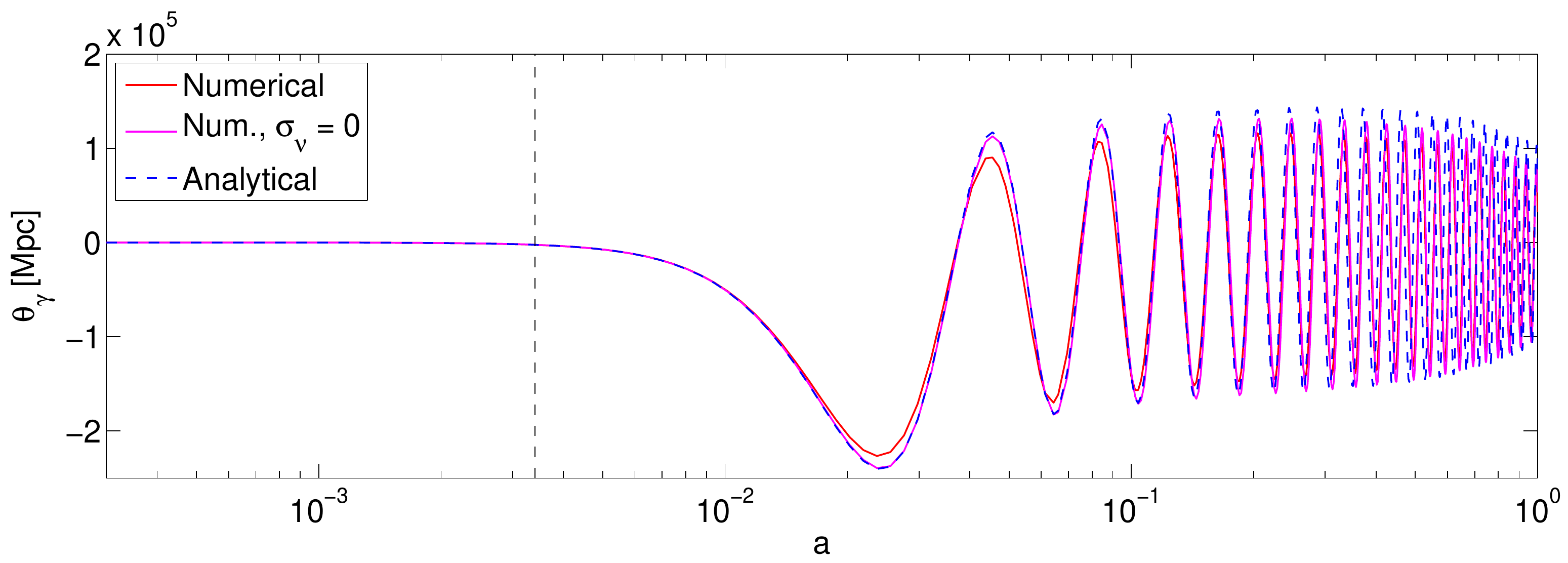}}
\end{minipage}
\begin{minipage}{1\columnwidth}
\centerline{\includegraphics[width=\linewidth]{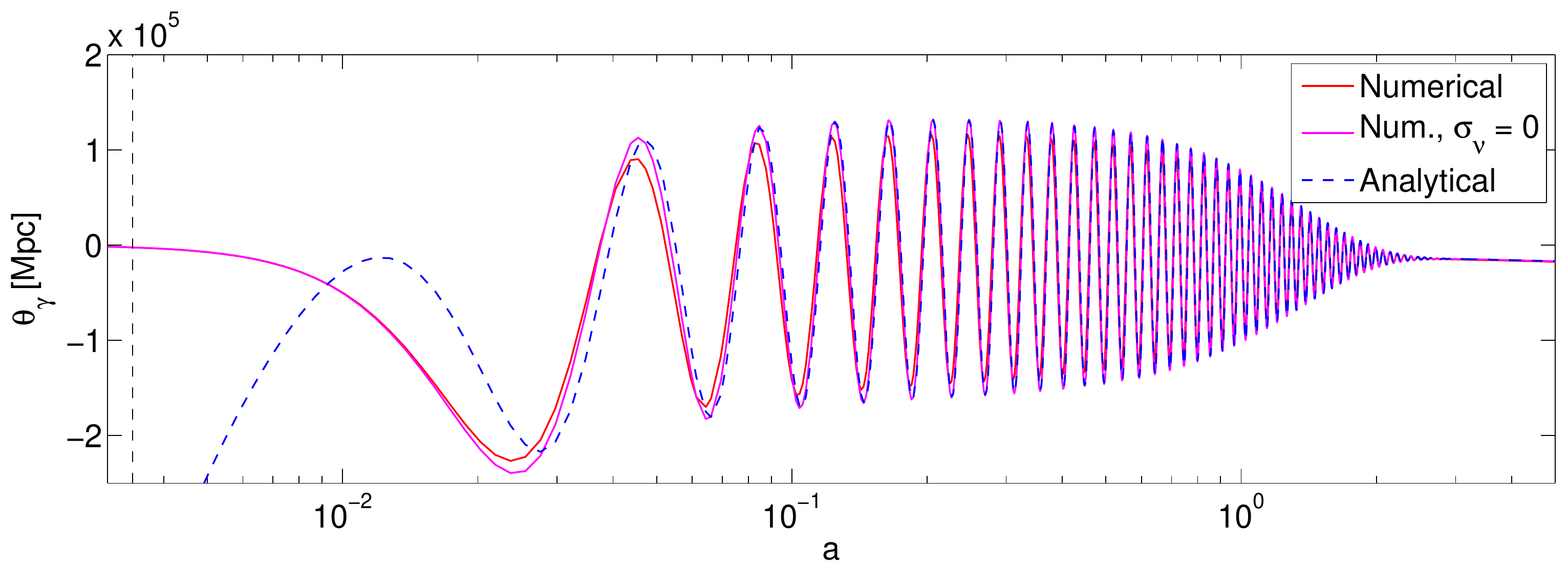}}
\end{minipage}
\caption{The radiation-domination \emph{(top panel)} and sub-Hubble \emph{(bottom panel)} solutions for $\theta_\gamma$ are compared with numerical simulations for a wavenumber $k=10.6 h/\text{Mpc}$. The slow mode is inferred from the exact hypergeometric solution.}
\label{fig:ug}
\end{figure}

The radiation-domination and sub-Hubble solutions for the plasma's velocity divergence are compared with numerical simulation in figure \ref{fig:ug}. Like the previous results, they both work very well inside their respective domains of validity. We observe the same offset in the amplitude of the fast mode, due to the effect of neglecting neutrino shear.

In contrast with density perturbations, the slow mode of the velocity divergence is not negligible and shifts the centre of oscillations. This is consistent with the findings of section~\ref{sec:summary}, where we showed that $\theta_\gamma^{\mr{slow}}$ was smaller than $ \theta_\gamma^{\mr{fast}}$ by only one power of $\theta$ instead of two. When diffusion damping becomes important, the fast mode decays and the slow mode gets dominant. Here again, the phase of the oscillations is well described by the analytical solution.

\subsection{Metric perturbations}

\begin{figure}[h!]
\begin{minipage}{0.5\linewidth}
\centerline{\includegraphics[scale=0.4]{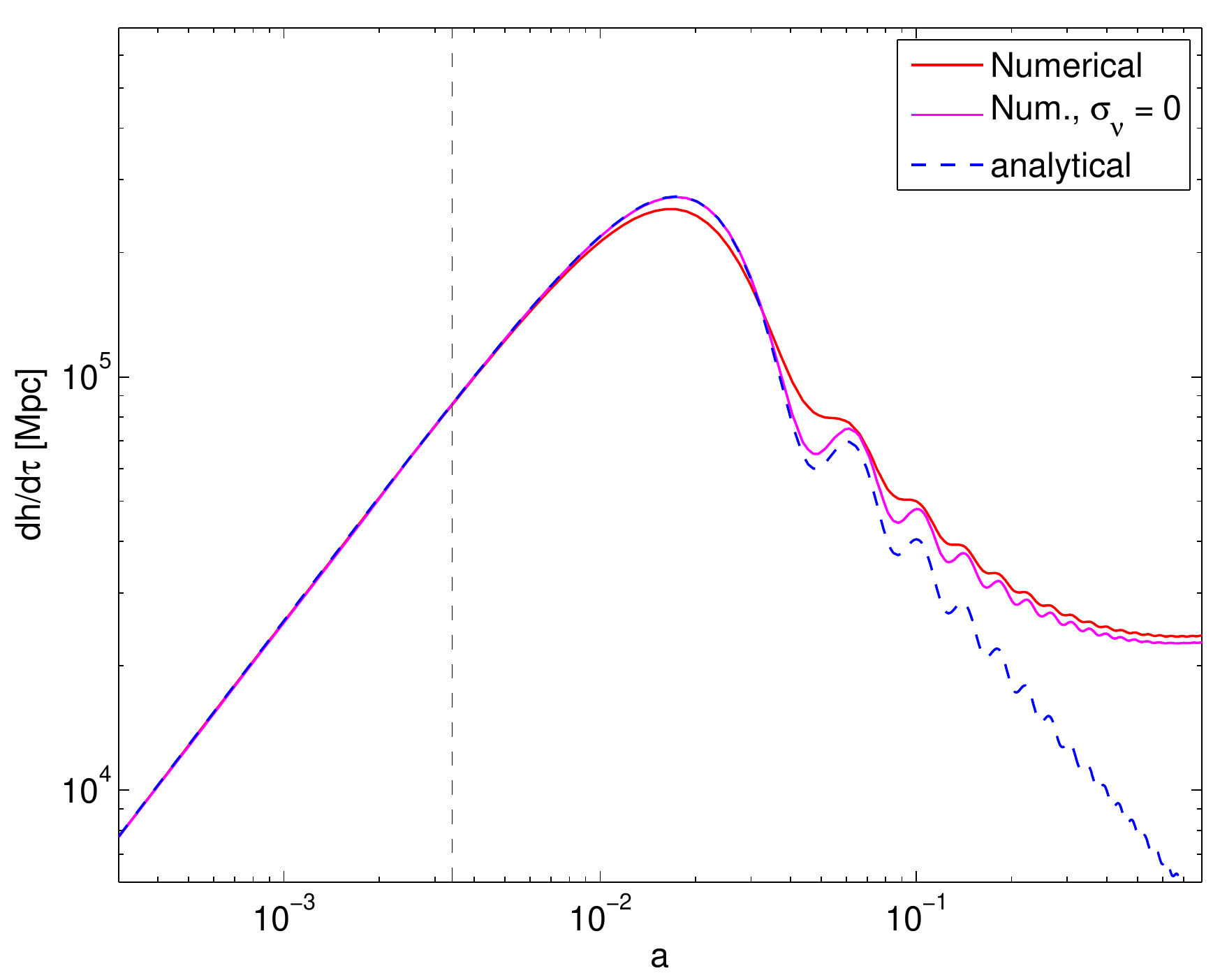}}
\end{minipage}
\hspace{0.15cm}
\begin{minipage}{0.5\linewidth}
\centerline{\includegraphics[scale=0.4]{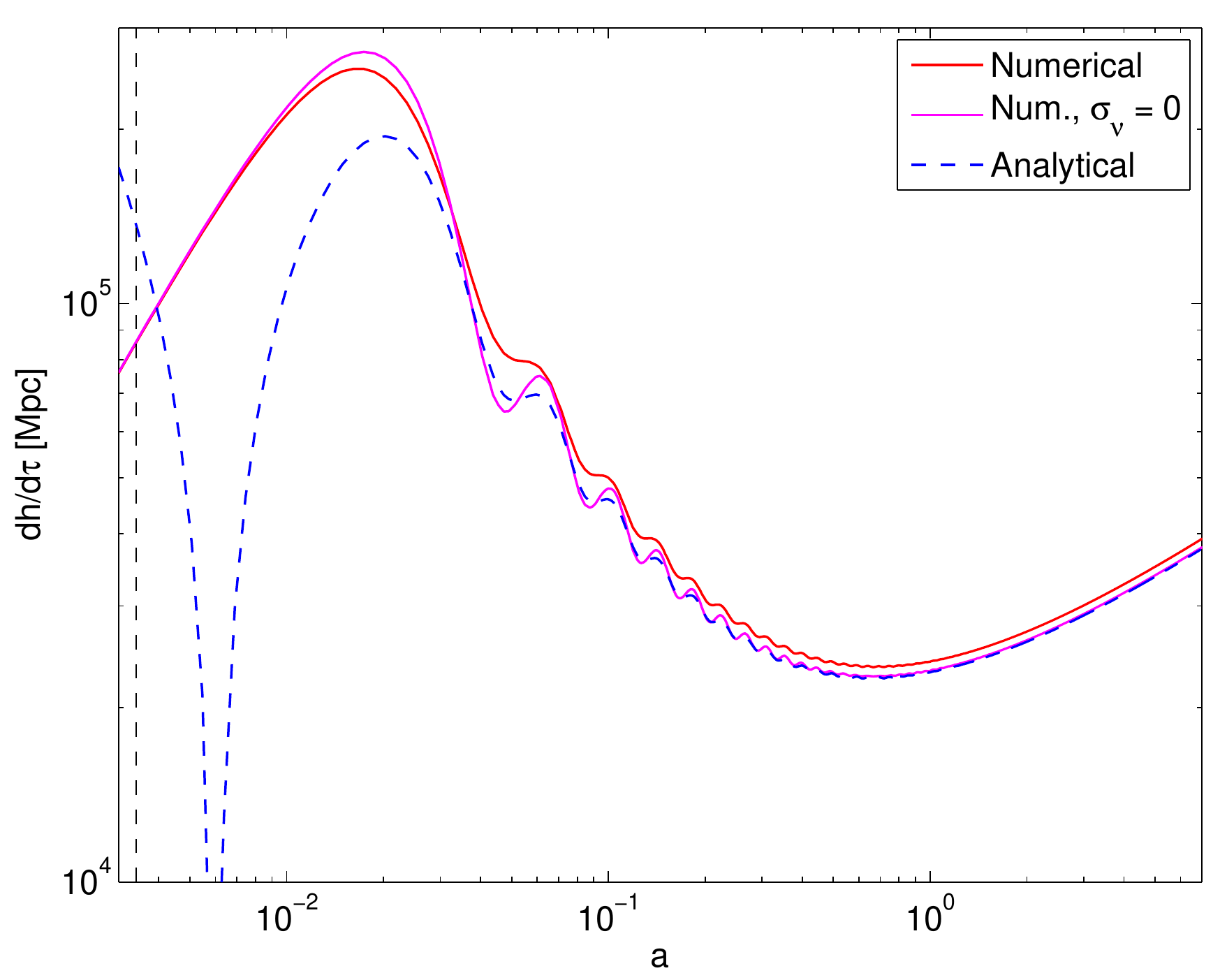}}
\end{minipage}
\caption{Comparison of numerical simulation and analytical solution for the metric perturbation $\dot{h}$ and a wavenumber $k=10.6 h/\text{Mpc}$.}
\label{fig:psi}
\end{figure}

Finally, the metric perturbation $\dot{h}$ is shown in figure \ref{fig:psi}. The small oscillations due to baryonic fluctuations are well captured by the analytic formula. Neglecting neutrino shear produces a small constant offset after Hubble crossing.

\section{CMB sensitivity to DM clustering}

Having explored analytical solutions on sub-Hubble scales, we can address the question: what is the CMB telling us exactly about dark matter?

The CMB is of course a very sensitive probe of the existence of a homogeneous background of dark matter, playing the role of a missing mass: it measures the total non relativistic matter density $\omega_m$ through the time of equality, affecting the height of the first few peaks, and the baryon density $\omega_b$ through the asymmetry between the first odd and even peaks. The difference is constrained to be $\omega_m-\omega_b=0.1198\pm 0.0026$ at the 68\% Confidence Level in the minimal $\Lambda$CDM model \cite{Ade:2013zuv,Planck:2013nga}. The CMB could also tell us something abut the annihilation or decay rate of DM, through its impact on the recombination and reionisation history (see e.g. \cite{Galli:2013dna} and references therein). On top of that, it is interesting to investigate whether the CMB provides any kind of constraint on the clustering properties of DM. 

\subsection{Contribution of fast and slow modes to the CMB}

The previous study showed that well inside the Hubble radius, there is an effective gravitational decoupling between the photon and dark matter components. One may try to infer from this observation that the CMB is decoupled from the evolution of CDM. If this was the case, the CMB would probe CDM only through its homogeneous density, not its gravitational effects. The argument could then be extended to show that alternative models coupled only gravitationally to other species (like, e.g., Warm Dark Matter, or self-interacting Dark Matter with no other interactions) cannot be probed with CMB observations.

This issue is not so obvious, because the CMB spectra do not depend only on the behaviour of the photon density perturbations $\delta_\gamma$. The source function which defines the spectrum of primary temperature anisotropies receives various contributions: the intrinsic temperature fluctuation, the Sachs-Wolfe (SW) correction, the Doppler effect, the early Integrated Sachs-Wolfe (ISW) effect, and a term related to the back-reaction of polarisation on temperature. All these terms relate to photon and baryon perturbations, excepted the SW and early ISW terms, related to metric perturbations. Secondary anisotropies induced e.g. by the late ISW effect or by CMB lensing also depend on metric fluctuations.

Instead, the polarisation source function only depends on photon perturbations. Polarisation is also affected by secondary effects, like CMB lensing, depending on metric fluctuations.

The CMB depends mainly on the source functions evaluated near the time of recombination (apart from the ISW and secondary contributions). For modes well-inside the Hubble radius at recombination, we could use the previous analytic solutions, and write each source function as the sum of a fast mode and a slow mode. Since fast modes are independent of the CDM evolution, while on the contrary slow modes are driven by CDM, it is important to know whether the source functions are dominated by the fast mode at recombination. If this is the case, the CMB should be insensitive to gravitational interactions with CDM, and more generally to the clustering properties of DM. 

\subsection{Probing DM clustering}

We expect from the results of the previous sections that photon and baryon perturbations are indeed dominated by fast modes at recombination, excepted on very small scales, for which they are washed out by Silk damping. Instead, metric fluctuations are dominated by slow modes on all scales. 

Before drawing conclusions from this observation, let us emphasise that this discussion refers only to cosmological models such that DM is coupled only gravitationally to other species. If this is not the case, i.e. if one allows for a non-negligible scattering rate between DM and photons, baryons, neutrinos, or possibly some other relativistic relics (see \cite{Wilkinson:2013kia,Dvorkin:2013cea,Serra:2009uu,Cyr-Racine:2013fsa} and references therein), then the DM component will couple also to fast modes. However, there are non-trivial models in which DM couples only gravitationally, while its density perturbations evolve very differently than in the plain CDM model. Indeed, DM clustering can be affected by free-streaming at large redshift (like for Warm Dark Matter) or by an internal pressure (like for self-interacting Dark Matter\footnote{We stress that we only refer here to models in which DM is self-interacting, but has no significant interactions with other particles. Our discussion cannot be extended to models where the internal pressure arises from a coupling with the photon-baryon plasma, with neutrinos or with other relics.}). Other effects can come into play, for instance in the case of Lorentz-violating Dark Matter \cite{Blas:2012vn}. In all these cases, density perturbations evolve in a generic way on large scales, while differences occur below a critical scale. Indeed, on very large (super-Hubble) scales, any DM component will obey 
\begin{equation}
\delta_c=\delta_b=\frac{3}{4} \delta_\gamma~,\label{eq:adiabatic}
\end{equation} 
unless it contains some entropy perturbations. Indeed, this relation should not be viewed as a consequence of the clustering properties of DM, but rather as an outcome of adiabatic initial conditions. Such conditions refer to a Universe perturbed by a single degree of freedom at initial time. When this is the case, Eq.~(\ref{eq:adiabatic}) can be derived from the equation of conservation of energy for each background species. Hence it is universal, and common to all DM models in which the background density scales like $\rho_c \propto a^{-3}$. The actual clustering properties of dark matter are encoded in quantities like its effective sound speed or viscosity coefficients, playing a role below a critical scale (that could be the free-streaming length or Jeans length). For plain CDM, this length is so small compared to observable CMB scales that it can be neglected. Since DM is a non-relativistic component, its effective sound speed cannot be as large as the speed of light, and the characteristic scale below which the clustering properties of different DM models can differ must be significantly smaller than the Hubble scale at a given time. Hence, if DM has non-trivial  clustering properties, we expect to see it only well-inside the Hubble radius.

Let us come back to the discussion of the sensitivity of the CMB to DM clustering. The SW and ISW terms are known to play an important role on large scales, typically those crossing the Hubble radius during matter domination (and hence not much smaller than the Hubble radius at recombination).  On those scales, metric perturbations are related to the density perturbations of all species. In this sense, the CMB temperature spectrum is not independent of DM perturbations on large angular scales (contributing to the SW plateau and to the first couple of acoustic peaks), but since they are universal, this does not provide a test of DM clustering properties.

For modes crossing the Hubble scales during radiation domination, and contributing to higher acoustic peaks, we know that metric fluctuations decay during radiation domination. They are much smaller than $\delta_\gamma$ at recombination, given by the fast mode, excepted on very small scales for which this fast mode is washed out by Silk damping. Hence slow modes can only play a role on such very small scales. As a consequence, the primary temperature and polarisation spectra should be independent of DM perturbations over a wide range of intermediate scales, ranging from the third acoustic peak up to some multipole deep in the damping scale. Therfore we expect the CMB to be a very bad probe of DM clustering properties.

Sensitivity to DM clustering could be restored through secondary effects. However the late ISW term is mainly affected by super-cluster scales; only DM models with some extreme pressure or velocity dispersion could affect such large scales. Similarly, CMB lensing mainly probes the scales corresponding to the maximum of the matter power spectrum $P(k)$, not very far from the scale $\lambda_\mathrm{eq}$ crossing the Hubble radius at equality, which is larger than the typical free-streaming or Jeans length of any non-cold DM candidate.

We conclude that the only region in which the CMB could be sensitive to DM properties is that of extremely small scales, well below the Silk damping horizon at recombination. There, the fast mode is washed out by the time of recombination, and slow modes driven by the evolution of CDM perturbations can emerge. Since on those scales the observed CMB spectra are dominated by foregrounds, we conclude that in practise, the CMB probes DM mainly through its background density, and provides no information on its clustering properties, as long as it couples only gravitationally to other species.

\subsection{Illustration with Warm Dark Matter}

We can demonstrate the validity of this argument by playing with Warm Dark Matter models. WDM is defined as a DM component with a velocity dispersion that cannot be neglected for the purpose of studying structure formation. Below its free-streaming scale, WDM does not cluster efficiently, simply because of diffusion processes. In pure WDM models, this effect induces an exponential cut-off in the matter spectrum~\cite{Colombi:1995ze}. However, the WDM phase-space distribution function can be such that the cut-off is very smooth, or looks like a step-like suppression~\cite{Boyarsky:2008mt}. While the second category of models is interesting phenomenologically, and is not strongly constrained by current cosmological data, we will illustrate our discussion with the most extreme deviation from CDM, namely a pure WDM model with a cut-off at a rather small $k$ value (even if ruled out by Lyman-$\alpha$ data). 

\begin{figure}[h!]
\begin{minipage}{0.5\linewidth}
\centerline{\includegraphics[scale=0.65]{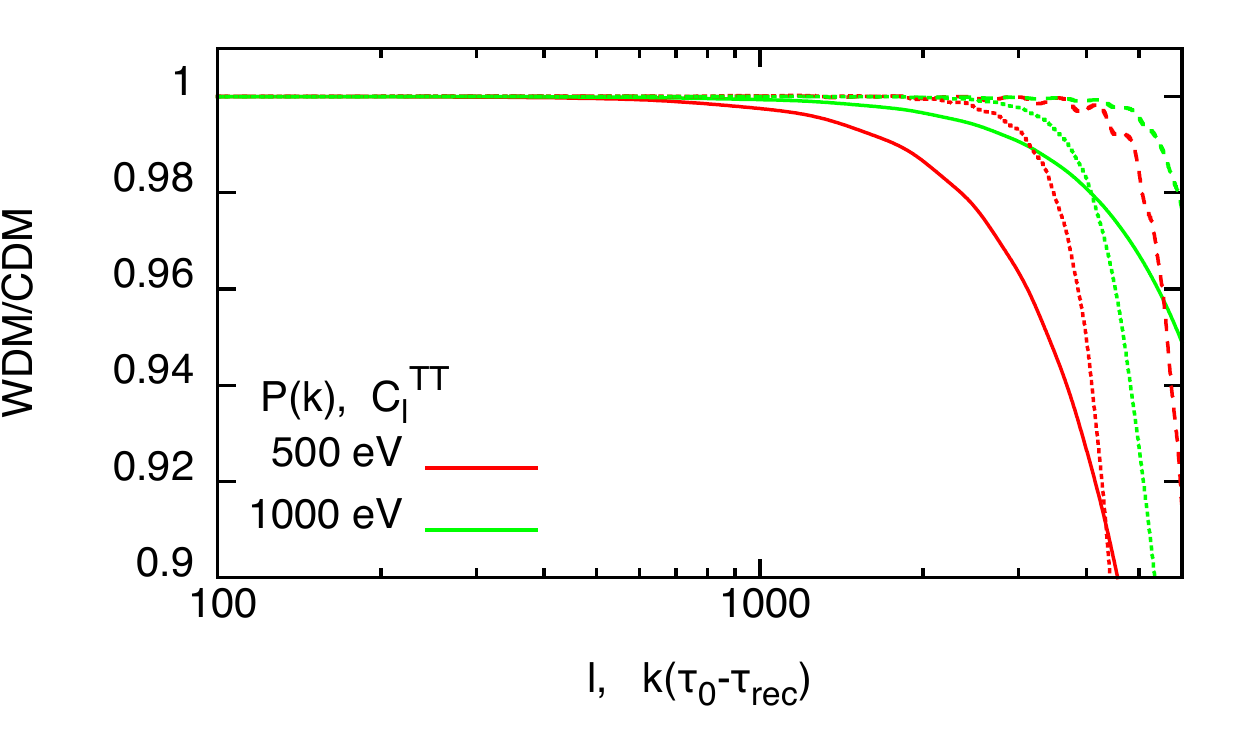}}
\end{minipage}
\hspace{-0.15cm}
\begin{minipage}{0.5\linewidth}
\centerline{\includegraphics[scale=0.65]{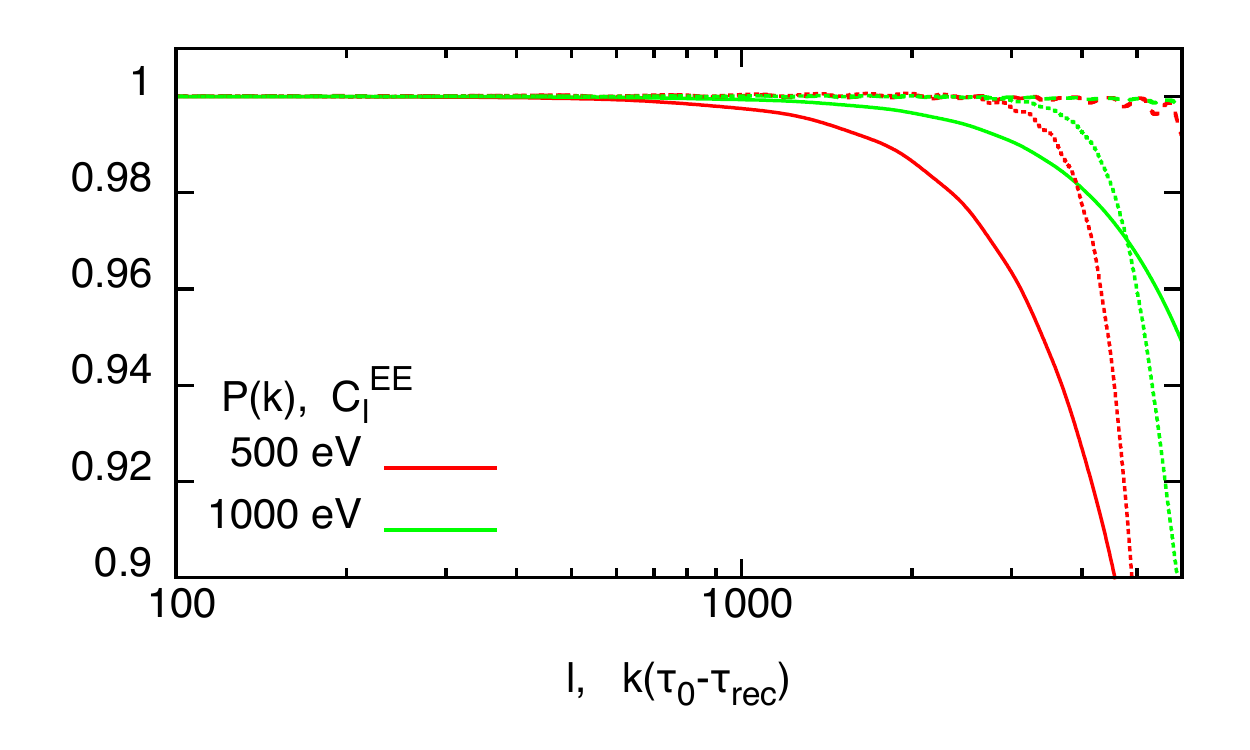}}
\end{minipage}
\caption{Power spectra of two pure WDM models, with a Dodelson-Widrow mass $m_1=500$~eV (red) or $m_2=1000$~eV (green), compare to those of a CDM model with the same cosmological parameters. In both plots, solid lines refer to matter power spectra $P(k)$, dashed lines to unlensed CMB spectra and dotted lines to lensed CMB spectra. However the dashed and dotted lines account for temperature on the left panel, and for polarisation on the right panel.}
\label{fig:wdm}
\end{figure}

In figure~\ref{fig:wdm}, we compare the matter power spectrum $P(k)$ and the unlensed/lensed temperature/polarisation power spectra $C_l^{TT}$, $C_l^{EE}$ derived: (i) from a CDM model, and (ii) from two pure WDM models, corresponding to the Dodelson-Widrow scenario~\cite{Colombi:1995ze}, and with two WDM masses $m_1=500$~eV and $m_2=1000$~eV.
In figure~\ref{fig:wdm}, the solid lines show the ratio of the respective matter power spectra $P(k)$. We see the free-streaming cut-off, which appears at a twice larger wavenumber $k_\mathrm{cut}$ for the mass $m_2$. Note that this cut-off is imprinted at a very high redshift $z_\mathrm{nr}$ (when WDM becomes non-relativistic), and keeps a fixed shape in comoving wavenumber space for $z<z_\mathrm{nr}$. Hence the figure shows the ratio of the power spectra calculated at any redshift $z<z_\mathrm{nr}$, including  the redshift of recombination, or $z=0$.

Instead of plotting the matter power spectrum as a function of $k$, we show it as a function of the dimensionless number $k(\tau_0-\tau_{rec})$: this corresponds to the multipole to which this comoving wavenumber contributes maximally at the time recombination. In the models of figure~\ref{fig:wdm}, the quantity $(\tau_0-\tau_{rec})$ is equal to $9530\,h^{-1}$Mpc. With such a rescaling, we can compare directly features in the matter power spectrum and in the primary CMB anisotropy spectra.

If the gravitational coupling between DM and photons played a role, we would expect the CMB temperature and polarisation spectra to be suppressed at the same scale $l_\mathrm{cut} = k_\mathrm{cut}(\tau_0-\tau_{rec})$ as the matter power spectrum. Indeed, on the scale where the cut-off is visible, the CDM model has a ratio of DM over photon density perturbations $(\rho_c \delta_c) / (\rho_\gamma \delta_\gamma)$ much larger than one during the end of radiation domination and throughout matter domination. Hence,  beyond $l_\mathrm{cut}$, one may naively expect that gravitational effects are more important in the CDM case than in the WDM case, and a feature should be visible in the CMB spectra.

But this is without counting on the effective gravitational decoupling discussed in the previous sections. We know that in the CDM case, DM perturbations are only relevant for slow modes, while the CMB is dominated by fast modes at least on intermediate scales. This conclusion can easily be extended to WDM. At very high redshift, when WDM is relativistic, it behaves like massless neutrinos, and it couples to fast modes. However the impact of WDM on fast modes is negligible, because WDM can only represent a tiny fraction of the radiation background. Indeed, any DM model reasonably fitting observations must have a background density scaling like $a^{-3}$ for an extended period of time before recombination\footnote{This would not be true for a very small mass (e.g. a Dodelson-Widrow mass $m \ll 100$~eV) for which dark matter would almost be hot.}. Extrapolating back in time, we see that when WDM becomes non-relativistic, it is much more diluted than ordinary neutrinos. For instance, in terms of effective neutrino number, the WDM component of the two models shown in figure~\ref{fig:wdm} contribute to radiation at early times respectively like $\Delta N_\mathrm{eff} = 0.0230$ (for $m_1$) or  $\Delta N_\mathrm{eff} = 0.0115$ (for $m_2$). Hence, the evolution of fast modes is affected by WDM by a negligible amount. After the non-relativistic transition, the WDM component obeys to the same equation of evolution as CDM, and can only contribute to slow modes.

Hence we expect the difference between the CDM and WDM models to be negligible in the CMB spectra, up to some very large multipole at which Silk damping washes out the fast mode of the photon/baryon perturbations, by such a large amount that the slow mode emerges. Looking at figure~\ref{fig:wdm}, we see that for our examples and for the unlensed temperature spectrum, this occurs at a multipole $l\sim 5000$, which is distinctively higher that the multipole corresponding to beginning of the cut-off in the matter spectrum ($l\sim 600$ or 1200). In the unlensed polarisation spectrum, the feature is pushed up to even higher $l$'s, because there is no SW effect for polarisation: hence a difference can appear only when the fast mode of photon/baryon perturbations become small with respect to their own slow mode, rather than to the slow mode of metric perturbations.

In the lensed spectra, the feature is moved to smaller $l$'s through smoothing effects (CMB lensing is known to correlate different multipoles and to make any feature smoother). Still, the lensed spectra are affected by WDM only above $l\sim 2500$, i.e. in a region hardly accessible to observations, due to foreground contamination. 

\vspace{0.5cm}

In summary of this section, we explained why the CMB has very little sensitivity to the clustering properties of dark matter, at least in the observable range of angular scales. We have illustrated our discussion with some particular example of pure WDM models. This clarifies the fact that the CMB has no sensitivity to WDM parameters, even when WDM induces a feature in the matter power spectrum on scales that can in principle be tested with CMB observations. This discussion could be extended to more general WDM models, or self-interacting DM models: the only important assumption is that DM couples only gravitationally to other species, in order to maintain the splitting between fast and slow modes described in the previous sections.

\section{Discussion and outlook}

We have presented the derivation of analytical sub-Hubble solutions for cosmological perturbations, valid until the time at which baryons decouple from photons. We followed the decomposition of the solutions into fast and slow modes first proposed by Weinberg \cite{Weinberg2002}. We worked with the standard notations of Ma \& Bertschinger \cite{MaBert}, in the synchronous gauge comoving with CDM. We found excellent agreement between these analytic solutions and numerical results from a Boltzmann code, up to differences that we can clearly attribute to the few approximations made in the analytical derivation, in particular, neglecting neutrino free-streaming.

The decomposition into fast and slow modes can be used to prove that there is an effective gravitational decoupling between CDM and the tightly coupled photon-baryon fluid. The fact that CDM does not feel the gravitational force from the photons (even for times and scales such that $\delta \rho_\gamma \ll \delta \rho_c$) comes from the fact that the wavefronts of photon density perturbations go across CDM potential wells over a timescale negligible with respect to the characteristic time of CDM clustering. Hence this force averages out to zero. The reciprocal effect comes from the fact that on sub-Hubble scales, photons experience pressure forces that are much larger than gravitational forces. The slow clustering of CDM shifts the zero-point of photon oscillations by a totally negligible amount compared to the amplitude of the oscillations. 

One striking consequence of this effective decoupling is that the evolution of CDM fluctuations can be studied using the M\'esz\'aros equation, or a variant of this equation accounting for baryonic corrections. This fact is very well-known since many decades, but the literature does not always provide a correct justification. The validity of the M\'esz\'aros  equation does not come from the fact that $\delta \rho_\gamma \ll \delta \rho_c$: during radiation domination, this inequality is not satisfied over an extended period of time after Hubble crossing. The correct explanation relies on the effective gravitational decoupling mechanism.

Another consequence is that the CMB is very weakly sensitive to the clustering properties of dark matter, as long as it couples only gravitationally to other species. Hence, CMB observations are inappropriate for discriminating among CDM, WDM or self-interacting DM models minimally coupled with other species. This does not come from the fact that the latter models would only affect the growth of matter perturbations on scales which are too small for impacting the CMB. Even in models in which the DM free-streaming length or Jeans length would be observable in principle, the CMB is unaffected, due to the effective decoupling. The sensitivity of the CMB to the clustering properties of DM is only restored on very small angular scales, for which fast modes are completely washed out by Silk damping. This sensitivity is further enhanced by gravitational lensing. Still, the signature of a possible non-standard DM clustering rate would only appear on scales at which the CMB signal is masked by foregrounds.

This set of analytic approximations could be improved, and used to speed-up Boltzmann codes. With further work, one could include in the calculation the effect of neutrino free-streaming and shear. The evolution of matter perturbations across the baryon drag time can be modelled like in Hu \& Sugiyama~\cite{HuSu96}. If analytic solutions can be brought to a sufficient level of accuracy, they could benefit to the efficiency of Boltzmann codes, and be substituted to the full integration over time of large wavelength perturbations. More specifically, Boltzmann codes sometimes need to calculate the matter transfer function $\delta_M(k,\tau)$ or the power spectrum $P(k,z)$ up to very large wavenumbers, either for the purpose of calculating lensed CMB spectra, or for deriving initial conditions for N-body simulations. Analytical solutions for $\delta_M$ could be implemented in the code, to make the calculation extremely fast for high $k$. Analytical solutions for photons could be used to derive the high-$l$ tail of the primary CMB spectra without performing a tedious integration over many oscillations in the photon-baryon fluid. We leave these possibilities for further investigation.

\appendix

\section{Radiation-dominated asymptotic solutions\label{sec:ARD}}

To get all asymptotic solutions in the radiation-dominated limit, one can plug the solution (\ref{eq:rdh}) and its time derivative into equations (\ref{RD4}, \ref{RD1}) to infer 
\be \delta_R (\tau) = - \frac{1}{6} \left( \tau\dot{h}(\tau) + \tau^2\ddot{h}(\tau) \right) = \frac{Nk^2}{18} \left[\frac{2}{\theta} \sin \theta - \left(1-\frac{2}{\theta^2}\right) \cos \theta  - \frac{2}{\theta^2} \right] \label{sol:GammaRD} \ee
and
\be \delta_c = - \frac{Nk^2}{6} \int_0^\theta \frac{1}{\theta^3}\left( \cos \theta + \theta \sin \theta - 1 - \frac{\theta^2}{2} \right)  d\theta ~.\label{sol:CDMRD} \ee 
We then calculate
\be \dot{\delta}_R = \frac{Nk^3}{18\sqrt{3}} \left[ \sin \theta \left(1 - \frac{4}{\theta^2} \right) + \frac{2 \cos \theta}{\theta}\left( 1 - \frac{2}{\theta^2}\right) + \frac{4}{\theta^3} \right]~.\ee
Therefore, the velocity perturbation reads
\be \theta_\gamma = - \left( \frac{\dot{h}}{2} + \frac{3}{4}\dot{\delta}_R \right) = - \frac{Nk^3}{24\sqrt{3}} \left( \sin \theta + \frac{2}{\theta}(\cos \theta -1 ) \right)~. \ee

\section{Matching the fast modes\label{sec:Match_fast}}

Having found the solution for $\delta_\gamma^{\mr{fast}}$ properly matched to the radiation-domination solution, equation (\ref{DeltaGammaFast}), one
can infer the other fast modes. To determine $\dot{h}^{\rm{fast}}$, we use Einstein equation (\ref{f_einstein}):
\be a \dot{h}^{\mr{fast}} = - 8 \pi G \int a^3 \rho_\gamma \left( 2 + R^{-1} \right) \delta_\gamma^{\mr{fast}} d\tau~.  \ee
Integrating the right-hand side, and ignoring the time dependence of all factors except the rapidly oscillating cosine, we find
\be
\dot{h}^{\mr{fast}} =\frac{4}{3\sqrt{3}} Nk \pi G a^2 \rho_\gamma (2+ R^{-1}) (1+R^{-1})^{\frac{1}{4}} \sin ( k r_s )~. \label{app:FinalHDotFast}
\ee
%
%
Using the Euler equations (\ref{f_continuity}) and proceeding in the same way, we find the fast modes for the cold dark matter density and the photon velocity: 
\begin{eqnarray}
\delta_c^{\rm{fast}} &=& \frac{2}{3} N \pi G a^2 \rho_\gamma (2+ R^{-1}) (1+R^{-1})^{\frac{3}{4}} \cos ( k r_s )~,\label{app:FinalDeltaCFast}\\
\theta_\gamma^{\mr{fast}} &=&   -  \frac{Nk^3}{24 \sqrt{3}} \frac{1}{(1+R^{-1})^{\frac{3}{4}}} \sin(kr_s)~. \label{app:FinalThetaGammaFast}
\end{eqnarray}

\section{Matching the slow modes\label{sec:Match_slow}}

Given the radiation-dominated solution (\ref{eq:RD_C_limit}), we can apply the matching condition (\ref{eq:matching}) to the solutions of the M\'esz\'aros equation, or of its generalisation to $\beta \neq 0$. In the M\'esz\'aros case, i.e. under the approximation $\beta = 0$, we calculate
\be \lim_{a \to 0} f_1(a)  =   1 \and \lim_{a \to 0} f_2(a) = - \log \left(\frac{a}{4}\right) - 3~.\ee
This matching condition fixes the coefficients of the linear combination  $\delta_c^{\rm{slow}} = \mathcal{A} f_1 + \mathcal{B} f_2$. The general slow solution for the CDM density perturbation then reads
\be \delta_c^{\mr{slow}} =\frac{Nk^2}{12}  \left[ \left(  \gamma + \log\left(\frac{2k}{\sqrt{2\pi G \rho_{eq}}}\right) -\frac{7}{2}\right)f_1 - f_2 \right] \qquad \qquad (\beta = 0)~. \label{app:FinalDeltaCSlow_simplified}\ee
To calculate the limit of the hypergeometric solution, we use the series expansion
\cite{Handbook}:
\be _2 F_1(a,b,a+b;z) = \frac{\Gamma(a+b)}{\Gamma(a)\Gamma(b)} \sum_{n=0}^{\infty} \frac{(a)_n(b)_n}{(n!)^2}\left[ 2\psi(n+1) - \psi(a+n)-\psi(b+n)-\log(1+z)\right](1-z)^n, \nonumber\ee
where $\Gamma$ and $\psi$ are the gamma and di-gamma function. In the limit $a\to 0$, $z \to 1$ and the $n=0$ term dominates:
\be \lim_{z \to 1} \, _2 F_1(a,b,a+b;z) = \frac{\Gamma(a+b)}{\Gamma(a)\Gamma(b)}\left[2\psi(1) - \psi(a)-\psi(b)-\log(1+z)\right]~. \ee
Therefore, the general slow solution for the CDM density perturbation with $\beta \neq 0$ reads
\be \delta_c^\mr{slow} = \frac{2}{3} \mathcal{C}(\alpha_{+}, \alpha_{-} ) \delta_c^{\mr{slow},-} + \frac{15}{4} \mathcal{C}(\alpha_{-}, \alpha_{+} ) \delta_c^{\mr{slow},+} , \label{app:FinalDeltaCSlow} \ee
where 
\begin{eqnarray}
\mathcal{C}(\alpha_{+},\alpha_{-}) =&& \frac{Nk^2}{12} \times \frac{\Gamma(\alpha_{-})\Gamma(\alpha_{-}+\frac{1}{2})}{\Gamma(2 \alpha_{-} + \frac{1}{2})}\frac{1}{\psi(\alpha_{+}) + \psi(\alpha_{+}+\frac{1}{2}) - \psi(\alpha_{-}) -\psi(\alpha_{-}+\frac{1}{2}) }\nonumber \\
&& \times \left\{2 \psi(1) -\psi(\alpha_+ ) - \psi(\alpha_+ + \frac{1}{2} ) + \log\left(\frac{2k}{\sqrt{2 \pi G \rho_{eq}}}\right) + \gamma -\frac{1}{2}-\log(4) \right\}~.\nonumber \\
&&
\end{eqnarray}
All other slow modes can easily be found by straightforward derivation. In the case $\beta = 0$ solutions, we calculate
\begin{eqnarray}
\frac{d f_1}{da} & = &\frac{3}{2}~, \qquad \qquad \frac{d f_2}{da} = \frac{3}{2}\log\left(\frac{\sqrt{1+a}+1}{\sqrt{1+a}-1}\right) - \frac{1+3a}{a\sqrt{1+a}}~, \\
\frac{d^2 f_1}{da^2}~,& = & 0 \qquad \qquad \frac{d^2 f_2}{da^2} = \frac{1}{a^2 (1+a)^{\frac{3}{2}}}~.  
\end{eqnarray}
This leads to the following set of simplified solutions:
\small
\begin{eqnarray}
&&-\frac{1}{2}\dot{h}^\mr{slow} = \frac{k^2}{4}\delta_\nu^\mr{slow}=\theta_\gamma^\mr{slow} = \theta_\nu^\mr{slow} = \nonumber \\
&& \frac{Nk^2}{12} \dot{a} \left[ \frac{3}{2}\( -\frac{7}{2} + \gamma + \log \( \frac{2k}{\sqrt{2\pi G \rho_{eq}}} \)  -  \log \(\frac{\sqrt{1+a}+1}{\sqrt{1+a}-1} \) \) + \frac{1+3a}{a\sqrt{1+a}} \right]~,\label{app:FinalOtherSlowModes_simplified}
\end{eqnarray}

\begin{center}
\begin{eqnarray}
\delta_\gamma^\mr{slow} &=& \frac{N}{3}\dot{a}^2(1+R^{-1}) \bigg\lbrace \left[ \frac{3}{2}\(-\frac{7}{2} + \gamma + \log \( \frac{2k}{\sqrt{2\pi G \rho_{eq}}} \)  - \log\left(\frac{\sqrt{1+a}+1}{\sqrt{1+a}-1}\right) \) + \frac{1+3a}{a\sqrt{1+a}} \right] \nonumber\\
&\times&  \(\frac{\ddot{a}}{\dot{a}^2} + \frac{1}{a(1+R)} \) - \frac{1}{a^2 (1+a)^{\frac{3}{2}}} \bigg\rbrace ~.\label{app:FinalDeltaGammaSlow_simplified}
\end{eqnarray}
\end{center}
\normalsize
For the exact solutions ($\beta \neq 0$), we use the formula 
\be \frac{\partial}{\partial z} \,_2F_1\left(a,b,c;z\right) = \frac{ab}{c} \,_2F_1\left(a+1,b+1,c+1;z\right)~. \nonumber\ee
This leads to the following set of exact solutions:

\small
\begin{eqnarray}
&&-\frac{1}{2}\dot{h}^\mr{slow} = \frac{k^2}{4}\delta_\nu^\mr{slow}=\theta_\gamma^\mr{slow} = \theta_\nu^\mr{slow} =  \nonumber \\
&&- \left[ \mathcal{C}(\alpha_{+}, \alpha_{-} )  (1+a)^{-(1+\alpha_-)} \alpha_- \dot{a} \,\, _2F_1 \(\alpha_- + \frac{1}{2}, \alpha_- + 1, 2 \alpha_- + \frac{1}{2};\frac{1}{1+a} \) + (\alpha_- \leftrightarrow \alpha_+ ) \right]~, \qquad \qquad\label{app:FinalOtherSlowModes}
\end{eqnarray}

\begin{eqnarray}
\delta_\gamma^\mr{slow} &=& \left[ \mathcal{C}(\alpha_{+}, \alpha_{-} ) \frac{1}{2a}(1+a)^{-(2+\alpha_-)} \alpha_- \bigg\lbrace (1+a)(2\alpha_- -1)\dot{a}^2 \,\, _2 F_1 \( \alpha_-, \alpha_- + \frac{1}{2}, 2 \alpha_- + \frac{1}{2}; \frac{1}{1+a}\) +  \right.  \nonumber \\
 &+& \left. \( (2+3a)\dot{a}^2 - 2a (1+a) \ddot{a} \) \,\, _2 F_1 \( \alpha_- + \frac{1}{2}, \alpha_- +1, 2 \alpha_- + \frac{1}{2}; \frac{1}{1+a}\) \bigg\rbrace + (\alpha_- \leftrightarrow \alpha_+ )  \right] . \label{app:FinalDeltaGammaSlow}
\end{eqnarray}

\normalsize

\bibliographystyle{utcaps}
\bibliography{article}

\providecommand{\href}[2]{#2}\begingroup\raggedright\begin{thebibliography}{10}

\bibitem{Planck}
{\bfseries Planck Collaboration} Collaboration, P.~Ade {\em et al.}, ``{Planck
  2013 results. I. Overview of products and scientific results},''
\href{http://arxiv.org/abs/1303.5062}{{\ttfamily arXiv:1303.5062
  [astro-ph.CO]}}.

\bibitem{Meszaros}
P.~Meszaros, ``{The behaviour of point masses in an expanding cosmological
  substratum},''
{\em Astron.Astrophys.} {\bfseries 37} (1974)  225--228.

\bibitem{Weinberg2002}
S.~Weinberg, ``{Cosmological fluctuations of short wavelength},''
  \href{http://dx.doi.org/10.1086/344441}{{\em Astrophys.J.} {\bfseries 581}
  (2002)  810--816},
\href{http://arxiv.org/abs/astro-ph/0207375}{{\ttfamily arXiv:astro-ph/0207375
  [astro-ph]}}.

\bibitem{CLASS1}
J.~Lesgourgues, ``{The Cosmic Linear Anisotropy Solving System (CLASS) I:
  Overview},''
\href{http://arxiv.org/abs/1104.2932}{{\ttfamily arXiv:1104.2932
  [astro-ph.IM]}}.

\bibitem{CLASS}
D.~Blas, J.~Lesgourgues, and T.~Tram, ``{The Cosmic Linear Anisotropy Solving
  System (CLASS) II: Approximation schemes},''
  \href{http://dx.doi.org/10.1088/1475-7516/2011/07/034}{{\em JCAP} {\bfseries
  1107} (2011)  034},
\href{http://arxiv.org/abs/1104.2933}{{\ttfamily arXiv:1104.2933
  [astro-ph.CO]}}.

\bibitem{Weinberg2008}
S.~Weinberg, {\em Cosmology}.
\newblock Oxford University Press, 2008.

\bibitem{MaBert}
C.-P. Ma and E.~Bertschinger, ``{Cosmological perturbation theory in the
  synchronous and conformal Newtonian gauges},''
  \href{http://dx.doi.org/10.1086/176550}{{\em Astrophys.J.} {\bfseries 455}
  (1995)  7--25},
\href{http://arxiv.org/abs/astro-ph/9506072}{{\ttfamily arXiv:astro-ph/9506072
  [astro-ph]}}.

\bibitem{GP75}
E.~J. Groth and P.~J.~E. Peebles, ``{Closed-form solutions for the evolution of
  density perturbations in some cosmological models},'' {\em Astron. \&
  Astrophys.} {\bfseries 41} (1975)  143--145.

\bibitem{HuSu96}
W.~Hu and N.~Sugiyama, ``{Small scale cosmological perturbations: An Analytic
  approach},'' \href{http://dx.doi.org/10.1086/177989}{{\em Astrophys.J.}
  {\bfseries 471} (1996)  542--570},
\href{http://arxiv.org/abs/astro-ph/9510117}{{\ttfamily arXiv:astro-ph/9510117
  [astro-ph]}}.

\bibitem{Lesgourgues:1519137}
J.~Lesgourgues, G.~Mangano, G.~Miele, and S.~Pastor, {\em {Neutrino
  cosmology}}.
\newblock Cambridge Univ. Press, Cambridge, 2013.

\bibitem{Ade:2013zuv}
{\bfseries Planck Collaboration} Collaboration, P.~Ade {\em et al.}, ``{Planck
  2013 results. XVI. Cosmological parameters},''
\href{http://arxiv.org/abs/1303.5076}{{\ttfamily arXiv:1303.5076
  [astro-ph.CO]}}.

\bibitem{Planck:2013nga}
{\bfseries Planck Collaboration} Collaboration, P.~Ade {\em et al.}, ``{Planck
  intermediate results. XVI. Profile likelihoods for cosmological
  parameters},''
\href{http://arxiv.org/abs/1311.1657}{{\ttfamily arXiv:1311.1657
  [astro-ph.CO]}}.

\bibitem{Galli:2013dna}
S.~Galli, T.~R. Slatyer, M.~Valdes, and F.~Iocco, ``{Systematic Uncertainties
  In Constraining Dark Matter Annihilation From The Cosmic Microwave
  Background},''
\href{http://arxiv.org/abs/1306.0563}{{\ttfamily arXiv:1306.0563
  [astro-ph.CO]}}.

\bibitem{Wilkinson:2013kia}
R.~J. Wilkinson, J.~Lesgourgues, and C.~Boehm, ``{Using the CMB angular power
  spectrum to study Dark Matter-photon interactions},''
\href{http://arxiv.org/abs/1309.7588}{{\ttfamily arXiv:1309.7588
  [astro-ph.CO]}}.

\bibitem{Dvorkin:2013cea}
C.~Dvorkin, K.~Bum, and M.~Kamionkowski, ``{Constraining Dark Matter-Baryon
  Scattering with Linear Cosmology},''
\href{http://arxiv.org/abs/1311.2937}{{\ttfamily arXiv:1311.2937
  [astro-ph.CO]}}.

\bibitem{Serra:2009uu}
P.~Serra, F.~Zalamea, A.~Cooray, G.~Mangano, and A.~Melchiorri, ``{Constraints
  on neutrino -- dark matter interactions from cosmic microwave background and
  large scale structure data},''
  \href{http://dx.doi.org/10.1103/PhysRevD.81.043507}{{\em Phys.Rev.}
  {\bfseries D81} (2010)  043507},
\href{http://arxiv.org/abs/0911.4411}{{\ttfamily arXiv:0911.4411
  [astro-ph.CO]}}.

\bibitem{Cyr-Racine:2013fsa}
F.-Y. Cyr-Racine, R.~de~Putter, A.~Raccanelli, and K.~Sigurdson, ``{Constraints
  on Large-Scale Dark Acoustic Oscillations from Cosmology},''
\href{http://arxiv.org/abs/1310.3278}{{\ttfamily arXiv:1310.3278
  [astro-ph.CO]}}.

\bibitem{Blas:2012vn}
D.~Blas, M.~M. Ivanov, and S.~Sibiryakov, ``{Testing Lorentz invariance of dark
  matter},'' \href{http://dx.doi.org/10.1088/1475-7516/2012/10/057}{{\em JCAP}
  {\bfseries 1210} (2012)  057},
\href{http://arxiv.org/abs/1209.0464}{{\ttfamily arXiv:1209.0464
  [astro-ph.CO]}}.

\bibitem{Colombi:1995ze}
S.~Colombi, S.~Dodelson, and L.~M. Widrow, ``{Large scale structure tests of
  warm dark matter},'' \href{http://dx.doi.org/10.1086/176788}{{\em
  Astrophys.J.} {\bfseries 458} (1996)  1},
\href{http://arxiv.org/abs/astro-ph/9505029}{{\ttfamily arXiv:astro-ph/9505029
  [astro-ph]}}.

\bibitem{Boyarsky:2008mt}
A.~Boyarsky, J.~Lesgourgues, O.~Ruchayskiy, and M.~Viel, ``{Realistic sterile
  neutrino dark matter with keV mass does not contradict cosmological
  bounds},'' \href{http://dx.doi.org/10.1103/PhysRevLett.102.201304}{{\em
  Phys.Rev.Lett.} {\bfseries 102} (2009)  201304},
\href{http://arxiv.org/abs/0812.3256}{{\ttfamily arXiv:0812.3256 [hep-ph]}}.

\bibitem{Handbook}
M.~Abramowitz and I.~Stegun, {\em Handbook of Mathematical Functions}.
\newblock Dover publications, New York, 1965.

\end{thebibliography}\endgroup
\end{document}